# gpu tracker: Python Package for Tracking and Profiling GPU and Other Hardware Utilization in Both Desktop and High-Performance Computing Environments

Erik D. Huckvale[1] and Hunter N.B. Moseley[1,2,3,4,5,6,*]

[1]Markey Cancer Center, University of Kentucky, Lexington, KY, USA

[2]Superfund Research Center, University of Kentucky, Lexington, KY, USA

[3]Center for Clinical and Translational Science, University of Kentucky, Lexington, KY, USA

[4]Department of Toxicology and Cancer Biology, University of Kentucky, Lexington, KY, USA

[5]Department of Molecular and Cellular Biochemistry, University of Kentucky, Lexington, KY, USA

[6]Institute for Biomedical Informatics, University of Kentucky, Lexington, KY, USA

[*]Corresponding Author: hunter.moseley@uky.edu

# Abstract

Over the lifetime of a computing task, determining the maximum usage of random-access memory (RAM) on both the motherboard and on a graphical processing unit (GPU), as well as the utilization percentage of the central processing unit (CPU) and GPU, can be extremely useful for troubleshooting points of failure as well as optimizing memory and processing unit utilization, especially within a high-performance computing (HPC) setting. While there are tools for tracking compute time, CPU utilization, and RAM, including by job management tools themselves, tracking of GPU usage, to our knowledge, does not currently have sufficient solutions, particularly in Unix/Linux operating systems. We present gpu-tracker, a multi-operating system Python package that tracks the computational resource usage of a task while running in the background, including the real compute time that the task takes to complete, its maximum RAM usage, the average and maximum percentage of CPU utilization, the maximum GPU RAM usage, and the average and maximum percentage of GPU utilization for both Nvidia and AMD GPUs. We demonstrate that gpu-tracker can seamlessly track computational resource usage with minimal overhead, both within desktop and HPC execution environments.

# 1 Introduction

Computation involves running computing tasks known as processes which last from their start to their finish for a duration known as compute time. Beyond time spent computing, these processes also consume other computational resources including random-access-memory (RAM) and at any point in time, the processes utilize a certain percentage of the central processing unit (CPU). There has been an increased demand for graphical processing units (GPU), which have their own RAM that is utilized by certain computing tasks, especially in the development of machine learning / artificial intelligence. Since computational resources are finite, it's imperative to estimate the resource usage of a computing task. This information aids in optimizing the implementation of computing tasks to make the most efficient use of these finite resources. It also informs users of how much computational resources to allocate for a task, particularly when submitting jobs to high-performance-computing (HPC) systems using job managers such as SLURM. If too many resources are allocated, it's wasteful, preventing other users from utilizing those resources for their own tasks, and it lowers one's priority in the job scheduling queue, necessitating that you wait longer for your job(s) to begin. And if too few resources are allocated, the jobs can crash prematurely, either due to reaching the specified time limit (wall time) or receiving out-of-memory errors. There are also different methods of accomplishing the same task, and the varying resource usage is valuable information when deciding which method is best, given the hardware constraints.

With the need to profile resource usage, tools have been developed for measuring compute time, the maximum amount of RAM used, and the CPU utilization over the lifetime of a process, including by job-scheduling tools themselves. However, tools for measuring maximum GPU usage are very few and insufficient for current use-cases, especially job-scheduling within HPC environments. For example, SLURM does not track GPU information at all and using SLURM to obtain compute time and RAM of a past job is not the most user-friendly experience, e.g. the availability of the information is ephemeral and the process ID of the past job must be known to obtain profiling information from older logs. Therefore, we introduce gpu-tracker, a tool that profiles compute time, RAM, CPU utilization percentage, GPU RAM, and GPU utilization percentage, for both Nvidia and AMD GPUs, all from a single application programming interface (API) and command-line interface (CLI). Users can integrate the profiling anywhere within their own Python scripts using the API and can profile an arbitrary shell command, Python or not, using the CLI. In this work, we

present:

• The gpu tracker software and its objected-oriented design.

• Description of gpu tracker's API and CLI.

• A tutorial demonstrating API and CLI use-cases.

• Profiling examples using the gpu tracker software.

• Related work with comparisons to gpu tracker.

• Conclusions.

# 2 Materials and Methods

All source code is available on our GitHub page at: https://github.com/MoseleyBioinformaticsLab/gpu_tracker. And gpu-tracker can be installed as a Python package via the Python Package Index at: https://pypi.org/project/gpu-tracker/.

## 2.1 Application Programming Interface (API)

### 2.1.1 Tracking

The primary module of the gpu-tracker API is called **`tracker`** and contains the **`Tracker`** class. Figure 1 shows a UML diagram of **`Tracker`** and related classes. A **`Tracker`** object is used to profile a block of Python code, either within a context manager, as indicated by its **`__enter__()`** and **`__exit__()`** methods (Sneeringer, 2015) or between explicit calls to its **`start()`** and **`stop()`** methods, collecting computational resource usage measurements including the maximum RAM usage, CPU utilization, maximum GPU RAM usage, GPU utilization, and the compute time (real time) through the duration that profiling occurs. While the default functionality is to track the computational resources used by the same process in which the **`Tracker`** itself is a part of, a **`Tracker`** object can also collect resource usage of an arbitrary process, specified by its **`int process_id`** parameter.

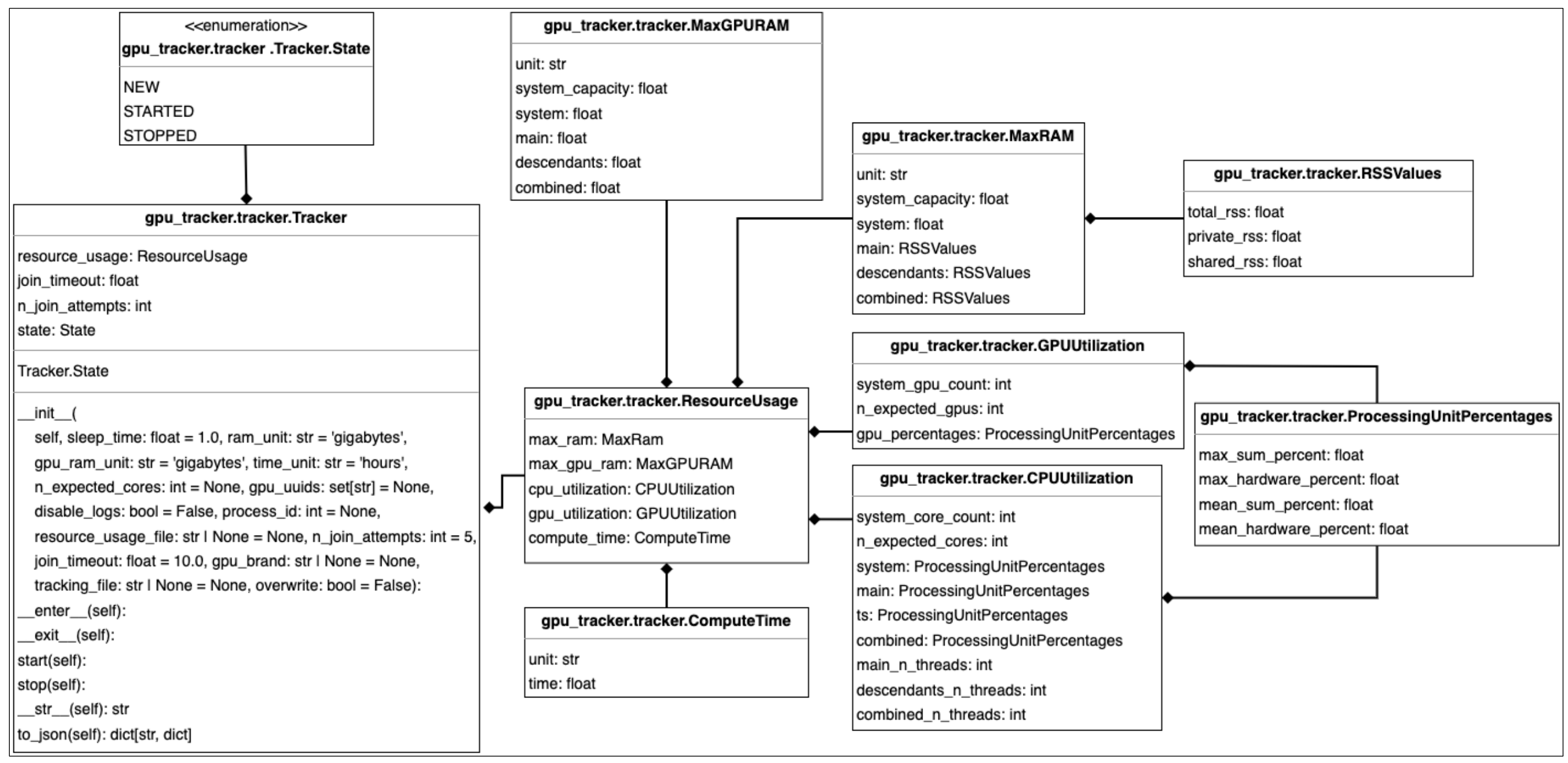

*Figure 1 - UML diagram of the Tracker class and the data classes it's composed of.*

The **`resource_usage`** attribute is a Python data class implemented using the **`@dataclass`** decorator from the **`dataclass`** module of the Python standard library. Introduced in Python 3.7, the **`dataclass`** module facilitates the implementation of data classes, which are designed to create data heavy objects with no (or very few) methods and are similar to a **`collections.namedtuple`** (Ramalho, 2022). The **`resource_usage`** has attributes which themselves are also data classes, including **`max_ram`**, **`max_gpu_ram`**, **`cpu_utilization`**, **`gpu_utilization`**, and **`compute_time`**. The data in these attributes is calculated after the profiling starts and is continuously updated until profiling stops. The **max_ram**, **max_gpu_ram**, and **compute_time** attributes contain the unit of measurement for the resource in addition to the measurements themselves. The **MaxRAM** data class is more verbose than the **MaxGPURAM** class to account for the complexities that arise from multiple processes sharing the same part of memory. The **float system_capacity** attribute is the RAM capacity of the entire operating system (OS), not to be confused with the **system** attribute which measures the amount of RAM that was actually used by the OS over the duration of tracking. The **MaxRAM** data class details the RAM usage of the main process as well as the summed RAM usage of any descendant processes it may have (i.e. child processes, grandchild processes etc.). The **combined** attribute is the sum of both the main process and its descendant processes. The measurements of the main, descendant, and combined processes are further divided within the **RSSValues** data class where the distinction between private resident-set-size (RSS) and shared RSS is made, where private RSS is the RAM that's exclusively used by a process and shared RSS is the RAM used by a process and at least one other process. The **float total_rss** attribute is the sum of **float private_rss** and **float shared_rss**. It's important to note that private RSS and shared RSS measurements are only available to gpu-tracker on Linux distributions, due the operating system limitations of the psutil package (Rodola, 2024). On a non-Linux OS, only the total RSS is available, which means the sum of the shared RSS of the descendant processes or combined processes could often lead to a gross overestimation. However, on Linux distributions, the shared RSS of a process is only counted once. However, this may lead to a slight underestimation due to limitations of Linux-reported shared RSS information; however, we consider this the best useful interpretation of the given information. The **MaxGPURAM** data class likewise divides its measurements in **main**, **descendants**, and **combined** attributes, but the distinction between shared and private memory is not made, the nvidia-smi package (Nvidia, 2024) and amd-smi package (AMD, 2024) only provide non-zero GPU RAM usage for one of a set of processes accessing the same shared GPU memory (Figure 1).

The **CPUUtilization** data class contains the **int system_core_count** attribute which is the total number of cores available to the OS as compared to the **int n_expected_cores** attribute which is the number of cores the user expects to actually be used by their process. This value defaults to the same value as **int system_core_count** and can be overridden by the **int n_expected_cores** parameter of the **Tracker** class's constructor. The **main** attribute of the **CPUUtilization** data class contains the CPU utilization percentage measurements of the main process provided by the **ProcessingUnitPercentages** data class. The **descendants** attribute is these percentages summed and the **combined** attributes is the sum of the main process and its descendant processes. Additionally, the `CPUUtilization` data class contains information related to threads including the maximum number of threads used by the main process and the summed number of threads in the descendant processes and combined within the `int main_n_threads`, `int descendants_n_threads`, and `int combined_n_threads` attributes. The `GPUUtilization` data class contains the `int system_gpu_count` attribute which is the total number of GPUs available to the OS and the `int n_expected_gpus` is the number of GPUs the user expects to actually be used by their process. This defaults to the same value as `int system_gpu_count`, but can be overridden to be the length of the `set[str] gpu_uuids` parameter of the `Tracker` constructor where the user specifies which GPUs they would like to collect utilization percentages for. The `ProcessingUnitPercentages` data class is used by both the `CPUUtilization` and `GPUUtilization` data classes, representing a has-a relationship. In fact, all of the class-level relationships in Figure 1 are has-a composition relationships representing composition rather than inheritance (i.e, is-a relationship). It has a `float max_sum_percent`, `float max_hardware_percent`, `float mean_sum_percent`, and `float mean_hardware_percent` attributes. The sum percent measures the summed percentage of all the processing units being used. The hardware percent measures the sum percent divided by the number of processing units being tracked (i.e. `int n_expected_cores` and/or `int n_expected_gpus`). Max measures the maximum percentage measured at any given time and the mean measures the average of the percentages measured over the lifetime of the process. Finally, the `ComputeTime` data class has the `float time` attribute which measures the real compute time (Figure 1).

The `Tracker` object can be represented as a string via its `__str__()` magic method, displaying all the measurements and any units of measurements within its data classes. The same information is displayed in a dictionary via the `to_json()` method. The user can specify the units through initialization parameters, namely `str ram_unit`, `str gpu_ram_unit`, and `str time_unit` (Figure 1). The valid `str` values for `ram_unit` and `gpu_ram_unit` are 'bytes', 'kilobytes', 'megabytes', 'gigabytes', and 'terabytes' (the default is 'gigabytes'). Valid `str` values for `time_unit` are 'seconds', 'minutes', 'hours', and 'days' (the default is 'hours'). The quantitative value calculated for each measurement will be scaled according to the unit chosen, e.g. 10 gigabytes of maximum GPU RAM used would result in a value of 10000.0 if 'megabytes' is chosen for `gpu_ram_unit`. An exception is raised if invalid units are provided, e.g. if 'milliseconds' is provided for time unit or any other invalid string.

Calling the `start()` method or entering the context manager will run a subprocess in the background that will collect resource usage stats at set intervals of time. The `float`

**sleep_time** parameter determines how frequently these stats are updated (a value of at least 0.1 seconds is required to ensure sufficient time to accurately measure CPU utilization). For example, if a value of 1.5 is chosen, the resource collection logic will be executed every 1.5 seconds. The resource collection loop determines the RAM and CPU utilization used at that interval of the process using the psutil Python package (Rodola, 2024) and the GPU RAM and GPU utilization of an Nvidia GPU is collected using the nvidia-smi package (Nvidia, 2024) (shell command in a subprocess) or of an AMD GPU using the amd-smi package (AMD, 2024). Which brand of GPU to profile defaults to whichever brand is detected in the system, checking Nvidia first, and can be overridden by the **str gpu_brand** parameter. If the RAM or GPU RAM is greater than the maximum so far, the **max_ram** and **max_gpu_ram** attributes are updated. The maximum utilization percentages are updated if higher than the maximum so far and average utilization is updated every time for both the **cpu_utilization** and **gpu_utilization** attributes. The **compute_time** attribute is additionally updated, calculated by taking the time point of the current resource collection interval and subtracting the time point at the initial call to **start()** or when the context manager was entered. After the profiling completes, the measurements of the **max_ram**, **max_gpu_ram**, **cpu_utilization**, **gpu_utilization**, and **compute_time** data classes remain constant.

While the **Tracker** object reports maximum and average measurements across the timepoints of a tracking session, measurements for each timepoint can be collected for more granular analyses by specifying the **str tracking_file** parameter, which can either be in CSV or SQLite (SQLite, n.d.) format. This file is also essential for sub-tracking as described in the next section. It additionally contains static data e.g. the number of cores or system RAM capacity as compared to the measurements that change at each time point e.g. the main process RAM or the GPU utilization. If the user created a tracking file from a previous tracking session, they'd like to create a new one of the same name, and they don't need the data from the previous session, they can overwrite the previous tracking file with a new one setting the **bool overwrite** parameter to **True** (default **False**).

The remaining constructor parameters relate to edge cases for ending the underlying subprocess when resource tracking completes or when a process's information fails to be collected at a given tracking interval. Resource tracking completion triggers the resource collection loop to stop and the underlying subprocess to end, i.e. the subprocess terminates and is closed. The **int n_join_attempts** parameter (default is 5) is the number of times the **Tracker** attempts to terminate its underlying subprocess. Under normal circumstances, the subprocess will end the first time and do so instantly. However, for exceptional circumstances, the **int join_timeout** parameter (default is 10 seconds) specifies how long to wait for the subprocess to terminate. If the timeout is reached, the **Tracker** object will attempt to terminate the subprocess again for up to a number of attempts equal to **n_join_attempts**, each time waiting for an amount of time equal to **join_timeout**. In the most exceptional case that the subprocess fails to end after all attempts, the subprocess is terminated by force and a warning is logged to the user, informing them of the failure. While there is not an option to disable these logged warning messages, there are also warnings logged during the tracking in case a process, usually a descendant process as compared to the main process, was detected initially at the beginning of the tracking interval but is closed before information about it can be measured due to race conditions. These warnings can be suppressed with the **bool disable_logs** parameter (default is **False**) (Figure 1).

### 2.1.2 Sub-tracking

While the **tracker** module involves obtaining overall measurements of a tracking session, the **sub_tracker** module involves tracking smaller blocks of code (which could be a particular function) by using the tracking file generated by the **tracking_file** parameter of the **Tracker** class. This begins with the **SubTracker** context manager class detailed in Figure 2. The **str code_block_name** parameter is the name of the code block being sub-tracked and defaults to the name of the file followed by a colon ':' and then the line number where the context manager is entered, i.e. where the code block begins. The line number can be replaced by specifying the **str code_block_attribute** parameter, but note that this parameter is not used if **code_block_name** is specified. A timestamp is logged at the beginning of the code block, i.e. entering the context manager and at the end of the code block, i.e. exiting the context manager. These timestamps are stored in a sub-tracking file specified by the **str sub_tracking_file** parameter. If the user created a sub-tracking file in a previous tracking session, but would like to create a new sub-tracking file of the same name and doesn't need the data from the previous session, the previous sub-tracking file can be overwritten by setting the **bool overwrite** parameter to **True** (default **False**). The **sub_tracker** module additionally contains a **sub_track** function decorator which can be used to sub-track a function and is equivalent to wrapping the **SubTracker** context manager around each call to said function.

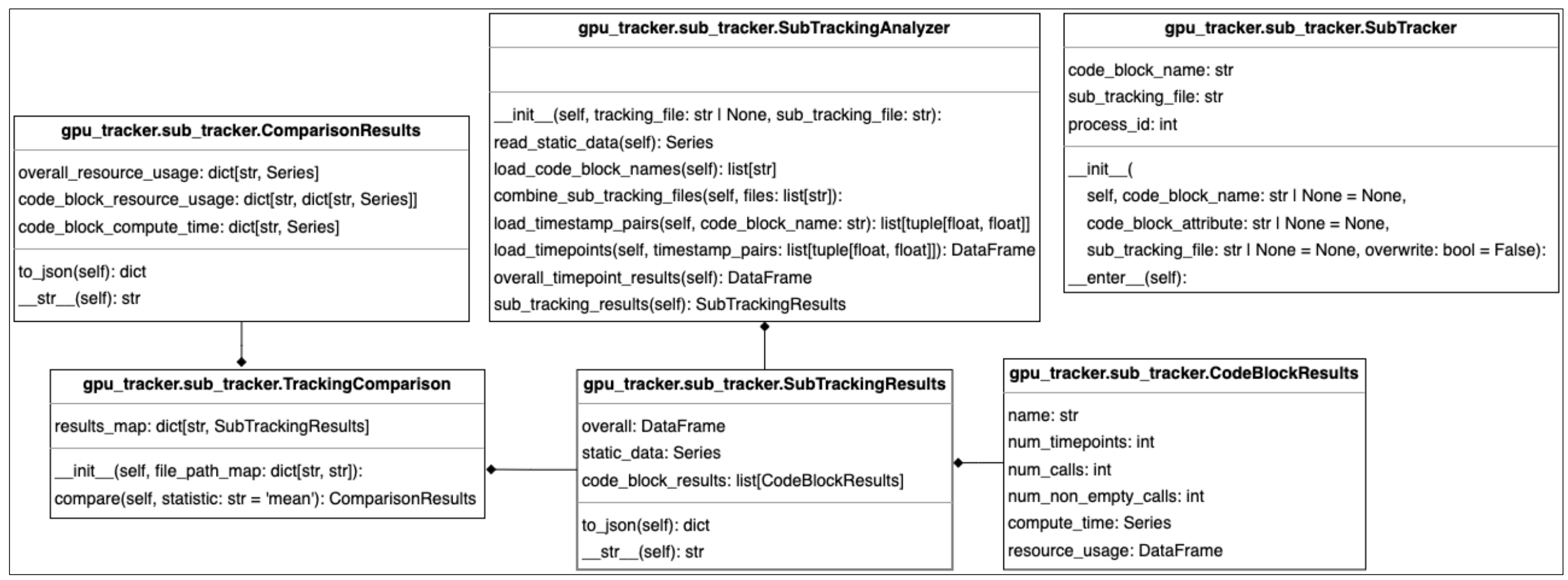

*Figure 2 - Classes within the sub_tracker module, including the SubTracker class which logs the timepoints of code blocks, the SubTrackingAnalyzer which uses that data to analyze the resource usage of each code block, and finally the TrackingComparison which compares the results of the SubTrackingAnalyzer across multiple tracking sessions.*

Using both the tracking file and sub-tracking file as specified by its **`str tracking_file`** and **`str sub_tracking_file`** constructor parameters, the **`SubTrackingAnalyzer`** class computes the resource usage of each code block that was sub-tracked within a tracking session. The **`SubTrackingAnalyzer`** stores this information in the **`SubTrackingResults`** data class returned by its **`sub_tracking_results`** method. All other methods of the **`SubTrackingAnalyzer`** are components of **`sub_tracking_results`** in case the user would like intermediary results leading up to the final **`SubTrackingResults`**, which contains the desired information in **`Series`** or **`DataFrame`** objects provided by the pandas Python package (The pandas development team, 2020). The **`overall`** attribute of the **`SubTrackingResults`** data class contains the overall resource usage across the entire tracking session, i.e. all the code that was tracked as compared to smaller blocks of code that were sub-tracked. The **`static_data`** attribute contains the static data stored in the tracking file. The information of primary interest is contained in the **`code_block_results`** attribute, which is a list of **`CodeBlockResults`** objects. The **`CodeBlockResults`** data class contains the resource usage information of the individual code blocks that were sub-tracked, including the **`str name`** attribute which is the name of the corresponding code block. The **`int num_timepoints`** attribute is the total number of timepoints from the tracking file that were measured within calls to the code block, i.e. that occurred at times within the start and stop timestamps (beginning of the code block and end of the code block) specified in the sub-tracking file. The **`int num_calls`** attribute is the number of times the sub-tracked code block was called (i.e. executed) within the tracked code block. Since it's possible that a code block can be called in such a short amount of time that no timepoints are recorded in the tracking file during that call, the **`int num_non_empty_calls`** attribute specifies the number of calls that had at least one timepoint measured during the time of the call. The **compute_time** attribute details the total time spent calling the code block (the sum of the time spent on all calls to the code block) as well as the minimum and maximum time spent calling a code block, the mean time spent across all code block calls, and the standard deviation. Finally, the **resource_usage** attribute details the minimum, maximum, mean, and standard deviation measurements for all computational resources across all timepoints. It is the same structure as the **overall** attribute of the **SubTrackingResults** class, but the measurements are for the particular code block.

If the **SubTrackingResults** objects of multiple different tracking sessions are saved in pickle format, the resource usage can be compared across these tracking sessions (e.g. for comparing how resource usage is affected by different implementations, different input data, etc.). The **TrackingComparison** object's constructor accepts a **dict[str, str] file_path_map** parameter, which maps the user given name of the tracking sessions to the file path containing the **SubTrackingResults** of the sessions. The **compare** method of the **TrackingComparison** class computes the **ComparisonResults,** which contains the resource usage statistics across the compared tracking sessions specified by the **file_path_map** constructor parameter. The **str statistic** parameter of the **compare** method allows the user to specify which statistic of the resource usages to compare across sessions, defaulting to the mean. While the **overall_resource_usage** attribute of the **ComparisonResults** class compares the overall resources used across the entire application, the **code_block_resource_usage** compares that for each code block, where exceptions are raised if the same code blocks were not tracked across sessions. The **code_block_compute_time** attribute compares the compute time statistic of each code block across sessions.

## 2.2 Command-Line Interface (CLI)

The gpu-tracker command-line interface (CLI), implemented using the docopt Python package (Keleshev, 2014), contains a command for tracking and three commands related to sub-tracking. The tracking command for tracking has only one required option, i.e. **--execute**, and simply wraps the **Tracker** class of the API. Under the hood, a **Tracker** object (see section 2.1 API) is used to profile the resource usage of a shell command executed in a subprocess (not to be confused with the subprocess used by the **Tracker** object to perform the tracking itself). Once the shell command's subprocess is started and its process ID is available, the process ID is passed into the **Tracker** constructor as the **int process_id** parameter so that the **Tracker** can monitor the resources of the subprocess. Several of the command-line options correspond to the configuration of the **Tracker** object, e.g. the **--st** option, which corresponds to the **sleep_time** parameter. The **-e** or **--execute** option specifies the command to run (along with the command's arguments) within quotes. After the command completes, the computational resource usage measurements are printed to the screen by default. The **--output** option can be set to a valid file path if the user wishes to store this output in a file rather than printing to the terminal. Additionally, the **--format** option controls the format of the output, both if printed and if stored in a file. The default format is plain text (output of the **__str__()** method of the Tracker object), but a JSON object containing the same information (output of the **to_json()** method of the **Tracker** object) is provided if **--format** is set to "json". The **ResourceUsage** object (**resource_usage** attribute of the Tracker class) can also be saved directly in "pickle" format.

The CLI also has the **sub-track** command under which are the **combine**, **analyze**, and **compare** sub-commands. The **combine** subcommand wraps the **combine_sub_tracking_files** method of the **SubTrackingAnalyzer** class in the **sub_tracker** module (see section 2.1 API), which accepts a list of sub-tracking files to combine into a single sub-tracking file. The **-p** option is used to specify this list of sub-tracking files to combine. If only one value is provided to the **-p** option, it's treated like a directory path and any files in that directory are treated as the list of sub-tracking files to combine. The **--stf** option specifies the path to the sub-tracking file that results from combining the files. The **analyze** sub-command wraps the **sub_tracking_results** method of the **SubTrackingAnalyzer** class. It also accepts a **--stf** option as well as a **--tf** option to specify the path of the sub-tracking file and tracking file to use to compute the results of the sub-tracking analysis. Like with the tracking command, the results of the analysis can be printed to the screen (default) or saved in a file specified by the **--output** option, with the format specified with the **--format** option: "text" (default), "json", or 'pickle' of the **SubTrackingResults** object (return value of the **sub_tracking_results** method). If at least two results from the **analyze** command are saved in pickle format, the results can be compared with the **compare** sub-command, which wraps the **compare** method of the **TrackingComparison** class. The **-m** option is used to specify the mapping from the name of each tracking session to its corresponding analysis results file such that the results within these multiple files can be compared. The **--stat** option specifies which summary statistic of the analyses to compare. The comparison results can likewise be printed to the screen (default) or stored in a file in text or JSON format or the **ComparisonResults** object (return value of the **compare** method) can be stored directly in pickle format. Figure 3 shows the remaining CLI options and their descriptions.

```
Usage:
    gpu-tracker -h | --help
    gpu-tracker -v | --version
    gpu-tracker --execute=<command> [--output=<output>] [--format=<format>] [--tconfig=<config-file>] [--st=<sleep-time>] [--ru=<ram-u
disable-logs] [--gb=<gpu-brand>] [--tf=<tracking-file>] [--overwrite]
    gpu-tracker sub-track combine --stf=<sub-track-file> -p <file-path>...
    gpu-tracker sub-track analyze --tf=<tracking-file> --stf=<sub-track-file> [--output=<output>] [--format=<format>]
    gpu-tracker sub-track compare [--output=<output>] [--format=<format>] [--cconfig=<config-file>] [-m <name>=<file-path>...] [--stat=<

Options:
    -h --help              Show this help message and exit.
    -v --version            Show package version and exit.
    -e --execute=<command>  The command to run along with its arguments all within quotes e.g. "ls -l -a".
    -o --output=<output>    File path to store the computational-resource-usage measurements in the case of tracking or the analysis rep
    -f --format=<format>    File format of the output. Either 'json', 'text', or 'pickle'. Defaults to 'text'.
    --tconfig=<config-file> JSON config file containing the key word arguments to the ``Tracker`` class (see API) to be optionally used in
override the corresponding arguments provided by the config file.
    --st=<sleep-time>       The number of seconds to sleep in between usage-collection iterations.
    --ru=<ram-unit>         One of 'bytes', 'kilobytes', 'megabytes', 'gigabytes', or 'terabytes'.
    --gru=<gpu-ram-unit>    One of 'bytes', 'kilobytes', 'megabytes', 'gigabytes', or 'terabytes'.
    --tu=<time-unit>        One of 'seconds', 'minutes', 'hours', or 'days'.
    --nec=<num-cores>       The number of cores expected to be used. Defaults to the number of cores in the entire operating system.
    --guuids=<gpu-uuids>    Comma separated list of the UUIDs of the GPUs for which to track utilization e.g. gpu-uuid1,gpu-uuid2,etc.
    --disable-logs          If set, warnings are suppressed during tracking. Otherwise, the Tracker logs warnings as usual.
    --gb=<gpu-brand>        The brand of GPU to profile. Valid values are nvidia and amd. Defaults to the brand of GPU detected in the s
    --tf=<tracking-file>    If specified, stores the individual resource usage measurements at each iteration. Valid file formats are CSV (.c
allows for more efficient querying.
    --overwrite             Whether to overwrite the tracking file if it already existed before the beginning of this tracking session. Do not set
    sub-track               Perform sub-tracking related commands.
    combine                 Combines multiple sub-tracking files into one. This is usually a result of sub-tracking a code block that is called in
    --stf=<sub-track-file>  The path to the sub-tracking file used to specify the timestamps of specific code-blocks. If not generated by th
called "data") where the headers are precisely process_id, code_block_name, position, and timestamp. The process_id is the ID of the
position is whether it is the start or the stopping point of the code block where 0 represents start and 1 represents stop. And timestamp
    -p <file-path>          Paths to the sub-tracking files to combine. Must all be the same file format and the same file format as the result
to a directory and all the files in this directory are combined.
    analyze                 Generate the sub-tracking analysis report using the tracking file and sub-tracking file for resource usage of specif
    compare                 Compares multiple tracking sessions to determine differences in computational resource usage by loading sub-t
running the ``sub-track analyze`` command and specifying a file path for ``--output`` and 'pickle' for the ``--format`` option. If code block
only overall results are compared.
    --cconfig=<config-file> JSON config file containing the ``file_path_map`` argument for the ``TrackerComparison`` class and ``statistic
``-m <name>=<path>`` and ``--stat=<statistic>`` commandline options respectively. If additional ``-m <name>=<path>`` options are add
the config file. If a ``--stat`` option is provided on the commandline, it will override the ``statistic`` in the config file.
    -m <name>=<file-path>   Mapping of tracking session names to the path of the file containing the sub-tracking results of said trackin
    --stat=<statistic>      The summary statistic of the measurements to compare. One of 'min', 'max', 'mean', or 'std'. Defaults to 'mean'.
```

*Figure 3 - The CLI help message shown by running 'gpu-tracker --help'.*

## 2.3 Example Profiling Analyses

The gpu-tracker package was used in the work of Huckvale and Moseley (Erik D Huckvale and Moseley, 2024c), i.e. the first profiling example, to profile the resource usage of their computational tasks. These tasks used HPC machines with a system capacity of 187 GB of RAM and 32 cores per node, with the CPUs being 'Intel® Xeon® Gold 6130 CPU@2.10GHz (Santa Clara, CA USA)'. Of these, no more than 17 gigabytes of RAM were allocated and no more than 16 cores were allocated. A GPU with 12 GB of GPU RAM was used, with the name of the GPU card being 'Tesla P100 PCIe 12 GB (Santa Clara, CA USA)'.

The hardware used for the second profiling example (Erik D Huckvale and Moseley, 2024b) included machines with up to 64 gigabytes (GB) of random-access memory (RAM) and central processing units (CPUs) ranging from 3.5 gigahertz (GHz) to 3.6 GHz, each with 6 hyperthreaded (HT) cores. The CPU chips included 'Intel(R) Core(TM) i7-6850K CPU@3.60GHz' and 'Intel(R) Core(TM) i7-5930K CPU@3.50GHz'. The graphic processing units (GPUs) used had 16 GB of GPU RAM, with the name of the GPU card being 'NVIDIA GeForce RTX 4080'. The GPUs were sourced from the Nvidia corporation in Santa Clara, California, USA.

The hardware used for the third profiling example (Erik D. Huckvale and Moseley, 2024) included machines with up to 2 terabytes (TB) of random-access memory (RAM) and central processing units (CPUs) of 3.8 gigahertz (GHz) of processing speed. The name of the CPU chip was "Intel(R) Xeon(R) Platinum 8480CL". The CPUs were sourced from the Intel corporation in Santa Clara, California, USA. The graphic processing units (GPUs) used had 81.56 gigabytes (GB) of GPU RAM, with the name of the GPU card being "NVIDIA H100 80GB HBM3". The GPUs were sourced from the Nvidia corporation in Santa Clara, California, USA.

# 3 Results

## 3.1 Tutorial

### 3.1.1 API

#### *3.1.1.1 Tracking*

##### 3.1.1.1.1 Basics

The **`gpu_tracker`** package provides the **`Tracker`** class which uses a subprocess to measure computational resource usage, namely the compute time, maximum CPU utilization, mean CPU utilization, maximum RAM used, maximum GPU utilization, mean GPU utilization, and maximum GPU RAM used. It supports both NVIDIA and AMD GPUs. The **`start()`** method starts this process which tracks usage in the background. The **`Tracker`** class can be used as a context manager. Upon entering the context, one can write the code for which resource usage is measured. The compute time will be the time from entering the context to exiting the context and the RAM, GPU RAM, CPU utilization, and GPU utilization quantities will be the respective computational resources used by the code that's within the context.

```
>>> import gpu_tracker as gput
>>> from example_module import example_function
>>> with gput.Tracker(n_expected_cores=1, sleep_time=0.1) as tracker:
>>>     example_function()
```

The **`Tracker`** class implements the **`__str__`** method so it can be printed as a string with the values and units of each computational resource formatted.

```
>>> print(tracker)

Max RAM:
   Unit: gigabytes
   System capacity: 67.254
   System: 4.307
   Main:
      Total RSS: 0.924
      Private RSS: 0.755
      Shared RSS: 0.171
   Descendants:
      Total RSS: 0.0
      Private RSS: 0.0
      Shared RSS: 0.0
   Combined:
      Total RSS: 0.924
      Private RSS: 0.755
      Shared RSS: 0.171
Max GPU RAM:
   Unit: gigabytes
   System capacity: 16.376
   System: 0.535
   Main: 0.314
   Descendants: 0.0
   Combined: 0.314
CPU utilization:
   System core count: 12
   Number of expected cores: 1
   System:
      Max sum percent: 222.6
      Max hardware percent: 18.55
      Mean sum percent: 149.285
```

```
    Mean hardware percent: 12.44
  Main:
    Max sum percent: 103.3
    Max hardware percent: 103.3
    Mean sum percent: 94.285
    Mean hardware percent: 94.285
  Descendants:
    Max sum percent: 0.0
    Max hardware percent: 0.0
    Mean sum percent: 0.0
    Mean hardware percent: 0.0
  Combined:
    Max sum percent: 103.3
    Max hardware percent: 103.3
    Mean sum percent: 94.285
    Mean hardware percent: 94.285
  Main number of threads: 15
  Descendants number of threads: 0
  Combined number of threads: 15
GPU utilization:
  System GPU count: 1
  Number of expected GPUs: 1
  GPU percentages:
    Max sum percent: 5.0
    Max hardware percent: 5.0
    Mean sum percent: 0.385
    Mean hardware percent: 0.385
Compute time:
  Unit: hours
  Time: 0.001
```

The output is organized by computational resource followed by information specific to that resource. The system capacity is a constant for the total RAM capacity across the entire operating system. There is a system capacity field both for RAM and GPU RAM. This is not to be confused with the system field, which measures the maximum RAM / GPU RAM (operating system wide) that was actually used over the duration of the computational-resource tracking. Both the RAM and GPU RAM have 3 additional fields, namely the usage of the main process itself followed by the summed usage of any descendant processes it may have (i.e. child processes, grandchild processes, etc.), and combined usage which is the sum of the main and its descendant processes. RAM is divided further to include the private RSS (RAM usage unique to the process), shared RSS (RAM that's shared by a process and at least one other process), and total RSS (the sum of private and shared RSS). The private and shared RSS values are only available on Linux distributions. For non-linux operating systems, the private and shared RSS will remain 0 and only the total RSS will be reported. Theoretically, the combined total RSS would never exceed the overall system RAM usage, but inaccuracies resulting from shared RSS can cause this to happen, especially for non-linux operating systems (see note below). The **`Tracker`** assumes that GPU memory is not shared across multiple processes and if it is, the reported GPU RAM of "descendant" and "combined" may be an overestimation.

The CPU utilization includes the system core count field which is the total number of cores available system-wide. Utilization is measured for the main process, its descendants, the main process and its descendants combined, and CPU utilization across the entire system. The sum percent is the sum of the percentages of all the cores being used. The hardware percent is the sum percent divided by the expected number of cores being used i.e. the optional **`n_expected_cores`** parameter (defaults to the number of cores in the entire system) for the main, descendants, and combined measurements. For the system measurements, hardware percent is divided by the total number of cores in the system regardless of the value of **`n_expected_cores`**. The max percent is the highest percentage detected through the duration of tracking while the mean percent is the average of all the percentages detected over that duration. The CPU utilization concludes with the maximum number of threads used at any time for the main process and the sum of the threads used across its descendant processes and combined.

The GPU utilization is similar to the CPU utilization, but rather than being based on utilization of processes, it can only measure the utilization percentages of the GPUs themselves, regardless of what processes are using them. To ameliorate this limitation, the optional **`gpu_uuids`** parameter can be set to specify which GPUs to measure utilization for (defaults to all the GPUs in the system). The system GPU count is the total number of GPUs in the system. The sum percent is the sum of all the percentages of these GPUs and the hardware percent is sum percent divided by the expected number of GPUs being used (i.e. **`len(gpu_uuids)`**). Likewise with CPU utilization, the max and mean of both the sum and hardware percentages are provided.

The compute time is the real time that the computational-resource tracking lasted (as compared to CPU time).

**NOTE**: *The keywords "descendants" and "combined" in the output above indicate a sum of the RSS used by multiple processes. It's important to keep in mind that on non-linux operating systems, this sum does not take into account shared memory but rather adds up the total RSS of all processes, which can lead to an overestimation. For Linux distributions, however, pieces of shared memory are only counted once.*

The **`Tracker`** can alternatively be used by explicitly calling its **`start()`** and **`stop()`** methods which behave the same as entering and exiting the context manager respectively.

```
>>> tracker = gput.Tracker()
>>> tracker.start()
>>> example_function()
>>> tracker.stop()
```

### 3.1.1.1.2 Arguments and Attributes

The units of the computational resources can be modified as desired. The following example measures the RAM in megabytes, the GPU RAM in megabytes, and the compute time in seconds.

```
>>> with gput.Tracker(ram_unit='megabytes', gpu_ram_unit='megabytes', time_unit='seconds', sleep_time=0.1) as tracker:
>>>     example_function()
>>> print(tracker)

Max RAM:
   Unit: megabytes
   System capacity: 67254.166
   System: 1984.791
   Main:
      Total RSS: 873.853
      Private RSS: 638.353
      Shared RSS: 235.68
   Descendants:
      Total RSS: 0.0
      Private RSS: 0.0
      Shared RSS: 0.0
   Combined:
      Total RSS: 873.853
      Private RSS: 638.353
      Shared RSS: 235.68
Max GPU RAM:
   Unit: megabytes
   System capacity: 16376.0
   System: 728.0
   Main: 506.0
   Descendants: 0.0
   Combined: 506.0
CPU utilization:
   System core count: 12
   Number of expected cores: 12
   System:
      Max sum percent: 161.6
      Max hardware percent: 13.467
      Mean sum percent: 145.517
      Mean hardware percent: 12.126
   Main:
      Max sum percent: 101.5
      Max hardware percent: 8.458
```

```
    Mean sum percent: 98.683
    Mean hardware percent: 8.224
  Descendants:
    Max sum percent: 0.0
    Max hardware percent: 0.0
    Mean sum percent: 0.0
    Mean hardware percent: 0.0
  Combined:
    Max sum percent: 101.5
    Max hardware percent: 8.458
    Mean sum percent: 98.683
    Mean hardware percent: 8.224
  Main number of threads: 15
  Descendants number of threads: 0
  Combined number of threads: 15
GPU utilization:
  System GPU count: 1
  Number of expected GPUs: 1
  GPU percentages:
    Max sum percent: 3.0
    Max hardware percent: 3.0
    Mean sum percent: 0.25
    Mean hardware percent: 0.25
Compute time:
  Unit: seconds
  Time: 2.729
```

The same information as the text format can be provided as a dictionary via the **`to_json()`** method of the **`Tracker`**.

```
>>> import json
>>> print(json.dumps(tracker.to_json(), indent=1))

{
 "max_ram": {
  "unit": "megabytes",
  "system_capacity": 67254.165504,
  "system": 1984.790528,
  "main": {
   "total_rss": 873.8529279999999,
   "private_rss": 638.353408,
   "shared_rss": 235.679744
  },
  "descendants": {
   "total_rss": 0.0,
   "private_rss": 0.0,
   "shared_rss": 0.0
  },
  "combined": {
   "total_rss": 873.8529279999999,
   "private_rss": 638.353408,
```

```
   "shared_rss": 235.679744
  }
 },
 "max_gpu_ram": {
  "unit": "megabytes",
  "system_capacity": 16376.0,
  "system": 728.0,
  "main": 506.0,
  "descendants": 0.0,
  "combined": 506.0
 },
 "cpu_utilization": {
  "system_core_count": 12,
  "n_expected_cores": 12,
  "system": {
   "max_sum_percent": 161.60000000000002,
   "max_hardware_percent": 13.466666666666669,
   "mean_sum_percent": 145.51666666666668,
   "mean_hardware_percent": 12.126388888888889
  },
  "main": {
   "max_sum_percent": 101.5,
   "max_hardware_percent": 8.458333333333334,
   "mean_sum_percent": 98.68333333333334,
   "mean_hardware_percent": 8.223611111111111
  },
  "descendants": {
   "max_sum_percent": 0.0,
   "max_hardware_percent": 0.0,
   "mean_sum_percent": 0.0,
   "mean_hardware_percent": 0.0
  },
  "combined": {
   "max_sum_percent": 101.5,
   "max_hardware_percent": 8.458333333333334,
   "mean_sum_percent": 98.68333333333334,
   "mean_hardware_percent": 8.223611111111111
  },
  "main_n_threads": 15,
  "descendants_n_threads": 0,
  "combined_n_threads": 15
 },
 "gpu_utilization": {
  "system_gpu_count": 1,
  "n_expected_gpus": 1,
  "gpu_percentages": {
   "max_sum_percent": 3.0,
   "max_hardware_percent": 3.0,
   "mean_sum_percent": 0.25,
   "mean_hardware_percent": 0.25
  }
 },
 "compute_time": {
```

```
  "unit": "seconds",
  "time": 2.728560209274292
 }
}
```

Using Python data classes, the **`Tracker`** class additionally has a **`resource_usage`** attribute containing fields that provide the usage information for each individual computational resource.

```
>>> tracker.resource_usage.max_ram

MaxRAM(unit='megabytes', system_capacity=67254.165504, system=1984.790528,
main=RSSValues(total_rss=873.8529279999999, private_rss=638.353408, shared_rss=235.679744),
descendants=RSSValues(total_rss=0.0, private_rss=0.0, shared_rss=0.0),
combined=RSSValues(total_rss=873.8529279999999, private_rss=638.353408, shared_rss=235.679744))

>>> tracker.resource_usage.max_ram.unit

'megabytes'

>>> tracker.resource_usage.max_ram.main

RSSValues(total_rss=873.8529279999999, private_rss=638.353408, shared_rss=235.679744)

>>> tracker.resource_usage.max_ram.main.total_rss

873.8529279999999

>>> tracker.resource_usage.max_gpu_ram

MaxGPURAM(unit='megabytes', system_capacity=16376.0, system=728.0, main=506.0, descendants=0.0,
combined=506.0)

>>> tracker.resource_usage.compute_time

ComputeTime(unit='seconds', time=2.728560209274292)
```

Below is an example of using a child process. Notice the descendants fields are now non-zero.

```
>>> import multiprocessing as mp
>>> ctx = mp.get_context(method='spawn')
>>> child_process = ctx.Process(target=example_function)
>>> with gput.Tracker(n_expected_cores=2, sleep_time=0.4) as tracker:
>>>     child_process.start()
```

```
>>>     example_function()
>>>     child_process.join()
>>> child_process.close()
>>> print(tracker)

Max RAM:
   Unit: gigabytes
   System capacity: 67.254
   System: 2.388
   Main:
      Total RSS: 0.849
      Private RSS: 0.528
      Shared RSS: 0.325
   Descendants:
      Total RSS: 0.845
      Private RSS: 0.734
      Shared RSS: 0.112
   Combined:
      Total RSS: 1.371
      Private RSS: 1.05
      Shared RSS: 0.325
Max GPU RAM:
   Unit: gigabytes
   System capacity: 16.376
   System: 1.236
   Main: 0.506
   Descendants: 0.506
   Combined: 1.012
CPU utilization:
   System core count: 12
   Number of expected cores: 2
   System:
      Max sum percent: 338.0
      Max hardware percent: 28.167
      Mean sum percent: 183.644
      Mean hardware percent: 15.304
   Main:
      Max sum percent: 101.0
      Max hardware percent: 50.5
      Mean sum percent: 60.178
      Mean hardware percent: 30.089
   Descendants:
      Max sum percent: 354.1
      Max hardware percent: 177.05
      Mean sum percent: 109.033
      Mean hardware percent: 54.517
   Combined:
      Max sum percent: 452.2
      Max hardware percent: 226.1
      Mean sum percent: 169.211
      Mean hardware percent: 84.606
```

```
  Main number of threads: 15
  Descendants number of threads: 13
  Combined number of threads: 28
GPU utilization:
  System GPU count: 1
  Number of expected GPUs: 1
  GPU percentages:
    Max sum percent: 5.0
    Max hardware percent: 5.0
    Mean sum percent: 0.556
    Mean hardware percent: 0.556
Compute time:
  Unit: hours
  Time: 0.001
```

Sometimes the code can fail. In order to collect the resource usage up to the point of failure, use a try/except block like so:

```
>>> try:
>>>     with gput.Tracker() as tracker:
>>>         example_function()
>>>         raise RuntimeError('AN ERROR')
>>> except Exception as error:
>>>     print(f'The following error occurred while tracking: {error}')
>>> finally:
>>>     print(tracker.resource_usage.max_gpu_ram.main)

The following error occurred while tracking: AN ERROR
0.506
```

If you do not catch the error in your code or if tracking otherwise is interrupted (e.g. you are debugging your code and you stop partway), the **`resource_usage`** attribute will not be set and that information will not be accessible in memory. In such a case, the **`resource_usage`** attribute will be stored in a hidden pickle file in the working directory with a randomly generated name. Its file path can be optionally overridden with the **`resource_usage_file`** parameter.

```
tracker = gput.Tracker(resource_usage_file='path/to/my-file.pkl')
```

While the **`Tracker`** class automatically detects which brand of GPU is installed (either NVIDIA or AMD), one can explicitly choose the GPU brand with the **`gpu_brand`** parameter

```
tracker = gput.Tracker(gpu_brand='nvidia')
```

While the **`Tracker`** by default stores aggregations of the computational resource usage across the timepoints, one can store the individual measured values at every timepoint in a file, either CSV or SQLite format, using the **`tracking_file`** parameter. **NOTE**: for the CSV format, the static data (e.g. RAM system capacity, number of cores in the OS, etc.) is stored on the first two rows with the headers on the first row followed by the static data on the second row. The headers of the timepoint data is on the third row followed by the timepoint data on the remaining rows. The SQLite file, however, stores the static data and timepoint data in different tables: "data" and "static_data" respectively.

```
>>> tracker = gput.Tracker(tracking_file='my-file.csv')
>>> tracker = gput.Tracker(tracking_file='my-file.sqlite')
```

#### *3.1.1.2 Sub-tracking*

##### 3.1.1.2.1 Logging Code Block Timestamps

While the **`Tracker`** object by itself can track a block of code, there are some cases where one might want to track one code block and a smaller code block within it or track multiple code blocks at a time without creating several tracking processes simultaneously, especially when tracking a code block that is executed within multi-processing or a code block that is executed several times. Similarly, one might want to track the resource usage of a particular function whenever it is called. Whether a function or some other specified code block, the **`SubTracker`** class can determine the computational resources used during the start times and stop times of a given code block. This includes the mean resources used during the instances the code block is executed, the mean time taken to complete the code block across these executions, the number of executions, etc. Sub-tracking uses the tracking file specified by the **`tracking_file`** parameter of the **`Tracker`** object alongside a sub-tracking file which contains the start and stop times of each code block one desires to sub-track. The sub-tracking file can be created in Python using the **`SubTracker`** class, a context manager around the desired code block. Setting the **`overwrite`** parameter (default **`False`**) of the **`Tracker`** and **`SubTracker`** to **`True`** overwrites the **`tracking_file`** or **`sub_tracking_file`** respectively if a file of that path already exists. Keep this parameter as **`False`** to avoid accidental loss of data.

```
>>> tracker = gput.Tracker(sleep_time=0.5, tracking_file='tracking.csv', overwrite=False)
>>> tracker.start()
>>> for _ in range(5):
>>>     with gput.SubTracker(code_block_name='my-code-block', sub_tracking_file='sub-tracking.csv',
overwrite=False):
```

```
>>>         example_function()
>>> @gput.sub_track(code_block_name='my-function', sub_tracking_file='sub-tracking.csv', overwrite=False)
>>> def my_function(*args, **kwargs):
>>>     example_function()

>>> for _ in range(3):
>>>     my_function()
>>> tracker.stop()
```

In the above example, a tracking session is initiated within the context of the **Tracker** object whose tracking file is ‘tracking.csv’. Then we have a for loop wherein a function is called 5 times. Other computation might be performed before or after this for loop, but if the computational resource usage of the contents of the for loop is of interest in particular, that code block can be sub-tracked by wrapping it within the context of the **SubTracker** object whose sub-tracking file is ‘sub-tracking.csv’. Alternatively, SQLite (**.sqlite**) files can be used to speed up querying in the case of very long tracking sessions. The name of the code block is ‘my-code-block’, given to distinguish it from other code blocks being sub-tracked.

If one wants to sub-track all calls to a particular function, the **sub_track** function decorator can be used instead of wrapping the function call with a **SubTracker** context every time it is called.

When sub-tracking a code block using the **SubTracker** context, the default **code_block_name** is the relative path of the Python file followed by a colon followed by the line number where the **SubTracker** context is initialized. When sub-tracking a function, the default **code_block_name** is the relative path of the Python file followed by a colon followed by the name of the function.

##### 3.1.1.2.2 Analysis

Once a tracking file and at least one sub-tracking file have been created, the results can be analyzed using the **SubTrackingAnalyzer** class, instantiated by passing in the path to the tracking file and the path to the sub-tracking file.

```
analyzer = gput.SubTrackingAnalyzer(tracking_file='tracking.csv', sub_tracking_file='sub-tracking.csv')
```

When sub-tracking a code block within a function that’s part of multi-processing (i.e. called within one of multiple sub-processes), the sub-tracking file must be unique to that process, which is why the default **sub_tracking_file** is the process ID followed by “.csv”. One way or another, a different sub-tracking file must be created per worker to prevent multiple processes from logging to the same file. The **SubTrackingAnalyzer** has a **combine_sub_tracking_files** method that can combine these multiple sub-tracking files into a single sub-tracking file whose path is specified by the **sub_tracking_file** parameter above. Once a sub-tracking file is created from a single process or combined from multiple, the results can be obtained via the **sub_tracking_results** method.

```
>>> results = analyzer.sub_tracking_results()
>>> type(results)

gpu_tracker.sub_tracker.SubTrackingResults
```

The **sub_tracking_results** method returns a **SubTrackingResults** object which contains summary statistics of the overall resource usage (all time points in the tracking file) and the per code block resource usage (the timepoints for executions of a code block i.e. the start/stop times) as **DataFrame** or **Series** objects from the **pandas** package.

```
>>> results.overall
```

| | min | max | mean | std |
|---|---|---|---|---|
| **main_ram** | 0.341217 | 0.920560 | 0.861921 | 0.100084 |
| **descendants_ram** | 0.000000 | 0.000000 | 0.000000 | 0.000000 |
| **combined_ram** | 0.341217 | 0.920560 | 0.861921 | 0.100084 |
| **system_ram** | 4.602618 | 5.701517 | 5.281926 | 0.220270 |
| **main_gpu_ram** | 0.000000 | 0.506000 | 0.448364 | 0.151267 |
| **descendants_gpu_ram** | 0.000000 | 0.000000 | 0.000000 | 0.000000 |

| combined_gpu_ram | 0.000000 | 0.506000 | 0.448364 | 0.151267 |
|---|---|---|---|---|
| system_gpu_ram | 0.215000 | 0.727000 | 0.668909 | 0.152657 |
| gpu_sum_utilization_percent | 0.000000 | 0.000000 | 0.000000 | 0.000000 |
| gpu_hardware_utilization_percent | 0.000000 | 0.000000 | 0.000000 | 0.000000 |
| main_n_threads | 12.000000 | 15.000000 | 14.757576 | 0.791766 |
| descendants_n_threads | 0.000000 | 0.000000 | 0.000000 | 0.000000 |
| combined_n_threads | 12.000000 | 15.000000 | 14.757576 | 0.791766 |
| cpu_system_sum_utilization_percent | 15.400000 | 138.400000 | 121.918182 | 19.484617 |
| cpu_system_hardware_utilization_percent | 1.283333 | 11.533333 | 10.159848 | 1.623718 |
| cpu_main_sum_utilization_percent | 91.400000 | 103.300000 | 99.060606 | 2.571228 |
| cpu_main_hardware_utilization_percent | 7.616667 | 8.608333 | 8.255051 | 0.214269 |
| cpu_descendants_sum_utilization_percent | 0.000000 | 0.000000 | 0.000000 | 0.000000 |
| cpu_descendants_hardware_utilization_percent | 0.000000 | 0.000000 | 0.000000 | 0.000000 |
| cpu_combined_sum_utilization_percent | 91.400000 | 103.300000 | 99.060606 | 2.571228 |
| cpu_combined_hardware_utilization_percent | 7.616667 | 8.608333 | 8.255051 | 0.214269 |

The **SubTrackingResults** class additionally contains the static data i.e. the information that remains constant throughout the tracking session.

```
>>> results.static_data

ram_unit                    gigabytes
gpu_ram_unit                gigabytes
time_unit                       hours
ram_system_capacity         67.254166
gpu_ram_system_capacity        16.376
system_core_count                  12
n_expected_cores                   12
system_gpu_count                    1
n_expected_gpus                     1
Name: 0, dtype: object
```

The **code_block_results** attribute of the **SubTrackingResults** class is a list of **CodeBlockResults** objects, containing the resource usage and compute time summary statistics. In this case, there are two **CodeBlockResults** objects in the list since there were two code blocks sub-tracked in this tracking session.

```
>>> my_code_block_results, my_function_results = results.code_block_results
>>> type(my_code_block_results)

gpu_tracker.sub_tracker.CodeBlockResults
```

The **compute_time** attribute of the **CodeBlockResults** class contains summary statistics for the time spent on the code block, where **total** is the total amount of time spent within the code block during the tracking session, **mean** is the average time taken on each execution of the code block, etc. The **resource_usage** attribute provides summary statistics for the computational resources used during executions of the code block i.e. within the start/stop times.

```
>>> my_code_block_results.compute_time

min       2.630907
max       2.869182
mean      2.685580
std       0.102789
total    13.427902
dtype: float64

>>> my_code_block_results.resource_usage
```

| | min | max | mean | std |
|---|---|---|---|---|
| **main_ram** | 0.341217 | 0.912278 | 0.846999 | 0.122948 |
| **descendants_ram** | 0.000000 | 0.000000 | 0.000000 | 0.000000 |
| **combined_ram** | 0.341217 | 0.912278 | 0.846999 | 0.122948 |
| **system_ram** | 4.602618 | 5.261357 | 5.170665 | 0.147118 |
| **main_gpu_ram** | 0.000000 | 0.506000 | 0.415429 | 0.182971 |
| **descendants_gpu_ram** | 0.000000 | 0.000000 | 0.000000 | 0.000000 |
| **combined_gpu_ram** | 0.000000 | 0.506000 | 0.415429 | 0.182971 |
| **system_gpu_ram** | 0.215000 | 0.727000 | 0.635714 | 0.184676 |
| **gpu_sum_utilization_percent** | 0.000000 | 0.000000 | 0.000000 | 0.000000 |
| **gpu_hardware_utilization_percent** | 0.000000 | 0.000000 | 0.000000 | 0.000000 |
| **main_n_threads** | 12.000000 | 15.000000 | 14.619048 | 0.973457 |
| **descendants_n_threads** | 0.000000 | 0.000000 | 0.000000 | 0.000000 |
| **combined_n_threads** | 12.000000 | 15.000000 | 14.619048 | 0.973457 |
| **cpu_system_sum_utilization_percent** | 15.400000 | 138.400000 | 120.142857 | 24.347907 |

| cpu_system_hardware_utilization_percent | 1.283333 | 11.533333 | 10.011905 | 2.028992 |
|---|---|---|---|---|
| **cpu_main_sum_utilization_percent** | 91.400000 | 103.300000 | 98.652381 | 2.733243 |
| **cpu_main_hardware_utilization_percent** | 7.616667 | 8.608333 | 8.221032 | 0.227770 |
| **cpu_descendants_sum_utilization_percent** | 0.000000 | 0.000000 | 0.000000 | 0.000000 |
| **cpu_descendants_hardware_utilization_percent** | 0.000000 | 0.000000 | 0.000000 | 0.000000 |
| **cpu_combined_sum_utilization_percent** | 91.400000 | 103.300000 | 98.652381 | 2.733243 |
| **cpu_combined_hardware_utilization_percent** | 7.616667 | 8.608333 | 8.221032 | 0.227770 |

Additionally, the **CodeBlockResults** class also has attributes for the name of the code block, the number of times it was executed (i.e. "called") during the tracking session, the number of executions that included at least one timepoint, and the total number of timepoints measured within all executions of the code block.

```
>>> my_code_block_results.name, my_code_block_results.num_calls,
my_code_block_results.num_non_empty_calls, my_code_block_results.num_timepoints

('my-code-block', 5, 5, 21)
```

The analysis results can also be printed in their entirety. Alternatively, the **to_json** method can provide this comprehensive information in JSON format.

```
>>> print(results)

Overall:
                          min       max      mean       std
  main_ram             0.341860  0.944374  0.856037  0.125014
  descendants_ram      0.000000  0.000000  0.000000  0.000000
  combined_ram         0.341860  0.944374  0.856037  0.125014
  system_ram           4.859711  5.553644  5.253445  0.134081
  main_gpu_ram         0.000000  0.506000  0.429920  0.170432
```

```
   descendants_gpu_ram                          0.000000    0.000000    0.000000   0.000000
   combined_gpu_ram                             0.000000    0.506000    0.429920   0.170432
   system_gpu_ram                               0.215000    0.727000    0.650320   0.172010
   gpu_sum_utilization_percent                  0.000000    3.000000    0.120000   0.600000
   gpu_hardware_utilization_percent             0.000000    3.000000    0.120000   0.600000
   main_n_threads                              12.000000   15.000000   14.720000   0.842615
   descendants_n_threads                        0.000000    0.000000    0.000000   0.000000
   combined_n_threads                          12.000000   15.000000   14.720000   0.842615
   cpu_system_sum_utilization_percent          11.900000  133.400000  119.212000  22.741909
   cpu_system_hardware_utilization_percent      0.991667   11.116667    9.934333   1.895159
   cpu_main_sum_utilization_percent            78.000000  103.200000   96.924000   6.390767
   cpu_main_hardware_utilization_percent        6.500000    8.600000    8.077000   0.532564
   cpu_descendants_sum_utilization_percent      0.000000    0.000000    0.000000   0.000000
   cpu_descendants_hardware_utilization_percent  0.000000    0.000000    0.000000   0.000000
   cpu_combined_sum_utilization_percent        78.000000  103.200000   96.924000   6.390767
   cpu_combined_hardware_utilization_percent    6.500000    8.600000    8.077000   0.532564
Static Data:
     ram_unit gpu_ram_unit time_unit ram_system_capacity gpu_ram_system_capacity system_core_count
n_expected_cores system_gpu_count n_expected_gpus
    gigabytes    gigabytes     hours          67.254166              16.376             12            12             1            1
Code Block Results:
   Name:               my-code-block
   Num Timepoints:     12
   Num Calls:          3
   Num Non Empty Calls: 3
   Compute Time:
             min       max      mean       std     total
        2.580433  2.789909  2.651185  0.120147  7.953554
   Resource Usage:
                                                   min         max        mean        std
        main_ram                                 0.341860    0.936559    0.808736   0.167663
        descendants_ram                          0.000000    0.000000    0.000000   0.000000
        combined_ram                             0.341860    0.936559    0.808736   0.167663
        system_ram                               4.859711    5.553644    5.231854   0.191567
        main_gpu_ram                             0.000000    0.506000    0.363500   0.225892
        descendants_gpu_ram                      0.000000    0.000000    0.000000   0.000000
        combined_gpu_ram                         0.000000    0.506000    0.363500   0.225892
        system_gpu_ram                           0.215000    0.727000    0.583250   0.228088
        gpu_sum_utilization_percent              0.000000    0.000000    0.000000   0.000000
        gpu_hardware_utilization_percent         0.000000    0.000000    0.000000   0.000000
        main_n_threads                          12.000000   15.000000   14.416667   1.164500
        descendants_n_threads                    0.000000    0.000000    0.000000   0.000000
        combined_n_threads                      12.000000   15.000000   14.416667   1.164500
        cpu_system_sum_utilization_percent      11.900000  130.800000  113.641667  32.352363
        cpu_system_hardware_utilization_percent  0.991667   10.900000    9.470139   2.696030
        cpu_main_sum_utilization_percent        79.600000  103.100000   96.583333   6.726587
        cpu_main_hardware_utilization_percent    6.633333    8.591667    8.048611   0.560549
        cpu_descendants_sum_utilization_percent  0.000000    0.000000    0.000000   0.000000
        cpu_descendants_hardware_utilization_percent  0.000000    0.000000    0.000000   0.000000
        cpu_combined_sum_utilization_percent    79.600000  103.100000   96.583333   6.726587
        cpu_combined_hardware_utilization_percent  6.633333    8.591667    8.048611   0.560549

   Name:               my-function
```

```
Num Timepoints:      12
Num Calls:           3
Num Non Empty Calls: 3
Compute Time:
        min       max      mean       std     total
   2.538011  2.577679  2.553176  0.021419  7.659528
Resource Usage:
                                                min         max        mean       std
    main_ram                                 0.864592    0.944374    0.896998  0.034505
    descendants_ram                          0.000000    0.000000    0.000000  0.000000
    combined_ram                             0.864592    0.944374    0.896998  0.034505
    system_ram                               5.203415    5.315219    5.271566  0.038751
    main_gpu_ram                             0.314000    0.506000    0.490000  0.055426
    descendants_gpu_ram                      0.000000    0.000000    0.000000  0.000000
    combined_gpu_ram                         0.314000    0.506000    0.490000  0.055426
    system_gpu_ram                           0.535000    0.727000    0.711000  0.055426
    gpu_sum_utilization_percent              0.000000    3.000000    0.250000  0.866025
    gpu_hardware_utilization_percent         0.000000    3.000000    0.250000  0.866025
    main_n_threads                          15.000000   15.000000   15.000000  0.000000
    descendants_n_threads                    0.000000    0.000000    0.000000  0.000000
    combined_n_threads                      15.000000   15.000000   15.000000  0.000000
    cpu_system_sum_utilization_percent     120.300000  133.400000  124.566667  4.001439
    cpu_system_hardware_utilization_percent 10.025000   11.116667   10.380556  0.333453
    cpu_main_sum_utilization_percent        94.700000  103.200000   98.841667  2.677332
    cpu_main_hardware_utilization_percent    7.891667    8.600000    8.236806  0.223111
    cpu_descendants_sum_utilization_percent  0.000000    0.000000    0.000000  0.000000
    cpu_descendants_hardware_utilization_percent  0.000000    0.000000    0.000000  0.000000
    cpu_combined_sum_utilization_percent    94.700000  103.200000   98.841667  2.677332
    cpu_combined_hardware_utilization_percent  7.891667    8.600000    8.236806  0.223111
```

#### 3.1.1.2.3 Comparison

The **`TrackingComparison`** class allows for comparing the resource usage of multiple tracking sessions, both the overall usage of the sessions and any code blocks that were sub-tracked. This is helpful if one wants to see how changes to the session might impact the computational efficiency of it, such as changes to implementation, input data, etc. To do this, the **`TrackingComparison`** takes a mapping of the given name of a tracking session to the file path where a **`SubTrackingResults`** object is stored in pickle format. Say we had two tracking sessions and we wanted to compare them. First, we store the **`results`** of the first tracking session in a pickle file. If we'd like to re-use the same names for the **`tracking_file`** and **`sub_tracking_file`** in the second tracking session, we can safely set the **`overwrite`** argument to **`True`** since their data has been saved in 'results.pkl'.

```
>>> import pickle as pkl
>>> import os
```

```
>>> with open('results.pkl', 'wb') as file:
>>>     pkl.dump(results, file)
```

Once we have the results of the first tracking session saved, we can start a new tracking session in another run of the program that we are profiling. Say we made some code changes and we want to compare the two implementations, we can populate a new **`tracking_file`** and **`sub_tracking_file`** with data from the new tracking session.

```
>>> import gpu_tracker as gput
>>> from example_module import example_function
>>> import pickle as pkl

>>> @gput.sub_track(code_block_name='my-function', sub_tracking_file='sub-tracking.csv', overwrite=True)
>>> def my_function(*args, **kwargs):
>>>     example_function()

>>> with gput.Tracker(sleep_time=0.5, tracking_file='tracking.csv', overwrite=True):
>>>     for _ in range(3):
>>>         with gput.SubTracker(code_block_name='my-code-block', sub_tracking_file='sub-tracking.csv',
overwrite=True):
>>>             example_function()
>>>         my_function()
>>> results2 = gput.SubTrackingAnalyzer(tracking_file='tracking.csv', sub_tracking_file='sub-
tracking.csv').sub_tracking_results()
>>> with open('results2.pkl', 'wb') as file:
>>>     pkl.dump(results2, file)
```

The first tracking session stored its results in 'results.pkl', while the second tracking session stored its results in 'results2.pkl'. Say we decided to call the first session 'A' and the second session 'B'. The **`TrackingComparison`** object would be initialized like so:

```
>>> comparison = gput.TrackingComparison(file_path_map={'A': 'results.pkl', 'B': 'results2.pkl'})
```

Once the **`TrackingComparison`** is created, its compare method generates the **`ComparisonResults`** object detailing the computational resource usage measured in one tracking session to that of the other tracking sessions. The **`statistic`** parameter determines which summary statistic of the measurements to compare, defaulting to 'mean'. In this example, we will compare the maximum measurements by setting **`statistic`** to 'max'.

```
>>> results = comparison.compare(statistic='max')
>>> type(results)

gpu_tracker.sub_tracker.ComparisonResults
```

The **`overall_resource_usage`** attribute of the **`ComparisonResults`** class is a dictionary mapping each measurement to a **`Series`** comparing that measurement across all timepoints in one tracking session to another.

```
>>> results.overall_resource_usage.keys()

dict_keys(['main_ram', 'descendants_ram', 'combined_ram', 'system_ram', 'main_gpu_ram', 'descendants_gpu_ram',
'combined_gpu_ram', 'system_gpu_ram', 'gpu_sum_utilization_percent', 'gpu_hardware_utilization_percent',
'main_n_threads', 'descendants_n_threads', 'combined_n_threads', 'cpu_system_sum_utilization_percent',
'cpu_system_hardware_utilization_percent', 'cpu_main_sum_utilization_percent',
'cpu_main_hardware_utilization_percent', 'cpu_descendants_sum_utilization_percent',
'cpu_descendants_hardware_utilization_percent', 'cpu_combined_sum_utilization_percent',
'cpu_combined_hardware_utilization_percent'])
```

For example, we can compare the overall maximum 'main_ram' of tracking session 'A' to tracking session 'B'.

```
>>> results.overall_resource_usage['main_ram']

A    0.920560
B    0.944374
dtype: float64
```

The **`code_block_resource_usage`** attribute is a dictionary that compares the same resource usage but for each code block rather than overall.

```
>>> results.code_block_resource_usage.keys()

dict_keys(['main_ram', 'descendants_ram', 'combined_ram', 'system_ram', 'main_gpu_ram', 'descendants_gpu_ram',
'combined_gpu_ram', 'system_gpu_ram', 'gpu_sum_utilization_percent', 'gpu_hardware_utilization_percent',
'main_n_threads', 'descendants_n_threads', 'combined_n_threads', 'cpu_system_sum_utilization_percent',
'cpu_system_hardware_utilization_percent', 'cpu_main_sum_utilization_percent',
'cpu_main_hardware_utilization_percent', 'cpu_descendants_sum_utilization_percent',
'cpu_descendants_hardware_utilization_percent', 'cpu_combined_sum_utilization_percent',
'cpu_combined_hardware_utilization_percent'])
```

Each measurement is a dictionary mapping each code block name to the resources used across tracking sessions in that code block.

```
>>> results.code_block_resource_usage['main_ram'].keys()

dict_keys(['my-code-block', 'my-function'])
```

For example, the maximum 'main_ram' used by 'my-code-block' in tracking session 'A' can be compared to that of tracking session 'B'.

```
>>> results.code_block_resource_usage['main_ram']['my-code-block']

A    0.912278
B    0.936559
dtype: float64
```

Finally the **`code_block_compute_time`** attribute is a dictionary that compares the compute time summary statistics for each code block and for each tracking session.

```
>>> results.code_block_compute_time.keys()

dict_keys(['my-code-block', 'my-function'])
```

For example, we can compare the maximum compute time of 'my-code-block' in tracking session 'A' to that of tracking session 'B'.

```
>>> results.code_block_compute_time['my-code-block']

B    2.789909
A    2.869182
dtype: float64
```

The comparison results can also be printed in their entirety. Alternatively, the **`to_json`** method can provide this comprehensive information in JSON format.

```
>>> print(results)

Overall Resource Usage:
    Main Ram:
                 A         B
         0.92056  0.944374
    Descendants Ram:
             A    B
         0.0  0.0
    Combined Ram:
                 A         B
         0.92056  0.944374
    System Ram:
```

```
          B         A
   5.553644  5.701517
Main Gpu Ram:
       A      B
   0.506  0.506
Descendants Gpu Ram:
     A    B
   0.0  0.0
Combined Gpu Ram:
       A      B
   0.506  0.506
System Gpu Ram:
       A      B
   0.727  0.727
Gpu Sum Utilization Percent:
     A    B
   0.0  3.0
Gpu Hardware Utilization Percent:
     A    B
   0.0  3.0
Main N Threads:
      A     B
   15.0  15.0
Descendants N Threads:
     A    B
   0.0  0.0
Combined N Threads:
      A     B
   15.0  15.0
Cpu System Sum Utilization Percent:
       B      A
   133.4  138.4
Cpu System Hardware Utilization Percent:
           B          A
   11.116667  11.533333
Cpu Main Sum Utilization Percent:
       B      A
   103.2  103.3
Cpu Main Hardware Utilization Percent:
     B         A
   8.6  8.608333
Cpu Descendants Sum Utilization Percent:
     A    B
   0.0  0.0
Cpu Descendants Hardware Utilization Percent:
     A    B
   0.0  0.0
Cpu Combined Sum Utilization Percent:
       B      A
   103.2  103.3
Cpu Combined Hardware Utilization Percent:
     B         A
   8.6  8.608333
```

```
Code Block Resource Usage:
    Main Ram:
        my-code-block:
                   A         B
            0.912278  0.936559
        my-function:
                  A         B
            0.92056  0.944374
    Descendants Ram:
        my-code-block:
              A    B
            0.0  0.0
        my-function:
              A    B
            0.0  0.0
    Combined Ram:
        my-code-block:
                   A         B
            0.912278  0.936559
        my-function:
                  A         B
            0.92056  0.944374
    System Ram:
        my-code-block:
                   A         B
            5.261357  5.553644
        my-function:
                   B         A
            5.315219  5.701517
    Main Gpu Ram:
        my-code-block:
                A      B
            0.506  0.506
        my-function:
                A      B
            0.506  0.506
    Descendants Gpu Ram:
        my-code-block:
              A    B
            0.0  0.0
        my-function:
              A    B
            0.0  0.0
    Combined Gpu Ram:
        my-code-block:
                A      B
            0.506  0.506
        my-function:
                A      B
            0.506  0.506
    System Gpu Ram:
        my-code-block:
                A      B
```

```
            0.727  0.727
    my-function:
                A      B
            0.727  0.727
Gpu Sum Utilization Percent:
    my-code-block:
              A    B
            0.0  0.0
    my-function:
              A    B
            0.0  3.0
Gpu Hardware Utilization Percent:
    my-code-block:
              A    B
            0.0  0.0
    my-function:
              A    B
            0.0  3.0
Main N Threads:
    my-code-block:
               A     B
            15.0  15.0
    my-function:
               A     B
            15.0  15.0
Descendants N Threads:
    my-code-block:
              A    B
            0.0  0.0
    my-function:
              A    B
            0.0  0.0
Combined N Threads:
    my-code-block:
               A     B
            15.0  15.0
    my-function:
               A     B
            15.0  15.0
Cpu System Sum Utilization Percent:
    my-code-block:
                B      A
            130.8  138.4
    my-function:
                A      B
            131.1  133.4
Cpu System Hardware Utilization Percent:
    my-code-block:
               B          A
            10.9  11.533333
    my-function:
                 A          B
            10.925  11.116667
```

```
    Cpu Main Sum Utilization Percent:
        my-code-block:
                  B      A
              103.1  103.3
        my-function:
                  A      B
              102.1  103.2
    Cpu Main Hardware Utilization Percent:
        my-code-block:
                     B         A
              8.591667  8.608333
        my-function:
                     A    B
              8.508333  8.6
    Cpu Descendants Sum Utilization Percent:
        my-code-block:
                A    B
              0.0  0.0
        my-function:
                A    B
              0.0  0.0
    Cpu Descendants Hardware Utilization Percent:
        my-code-block:
                A    B
              0.0  0.0
        my-function:
                A    B
              0.0  0.0
    Cpu Combined Sum Utilization Percent:
        my-code-block:
                  B      A
              103.1  103.3
        my-function:
                  A      B
              102.1  103.2
    Cpu Combined Hardware Utilization Percent:
        my-code-block:
                     B         A
              8.591667  8.608333
        my-function:
                     A    B
              8.508333  8.6
Code Block Compute Time:
    my-code-block:
                 B         A
          2.789909  2.869182
    my-function:
                 A         B
          2.570437  2.577679
```

Notice that the summary statistic values for each computational resource are ordered smallest to largest, which means the order the sessions changes for each computational resource, i.e. some are ordered “A B” and others are ordered “B A” in this example.

### 3.1.2 CLI

#### *3.1.2.1 Tracking*

##### 3.1.2.1.1 Basics

The **`gpu-tracker`** package also comes with a command-line interface that can track the computational-resource-usage of any shell command, not just Python code. Entering **`gpu-tracker -h`** in a shell will show the help message.

```
$ gpu-tracker -h

Tracks the computational resource usage (RAM, GPU RAM, CPU utilization, GPU utilization, and compute time) of a
process corresponding to a given shell command.

Usage:
    gpu-tracker -h | --help
    gpu-tracker -v | --version
    gpu-tracker --execute=<command> [--output=<output>] [--format=<format>] [--tconfig=<config-file>] [--st=<sleep-
time>] [--ru=<ram-unit>] [--gru=<gpu-ram-unit>] [--tu=<time-unit>] [--nec=<num-cores>] [--guuids=<gpu-uuids>] [--
disable-logs] [--gb=<gpu-brand>] [--tf=<tracking-file>] [--overwrite]
    gpu-tracker sub-track combine --stf=<sub-track-file> [-p <file-path>]...
    gpu-tracker sub-track analyze --tf=<tracking-file> --stf=<sub-track-file> [--output=<output>] [--format=<format>]
    gpu-tracker sub-track compare [--output=<output>] [--format=<format>] [--cconfig=<config-file>] [-m <name>=<file-
path>...] [--stat=<statistic>]

Options:
    -h --help               Show this help message and exit.
    -v --version            Show package version and exit.
    -e --execute=<command>  The command to run along with its arguments all within quotes e.g. "ls -l -a".
    -o --output=<output>    File path to store the computational-resource-usage measurements in the case of tracking
or the analysis report in the case of sub-tracking. If not set, prints to the screen.
    -f --format=<format>    File format of the output. Either 'json', 'text', or 'pickle'. Defaults to 'text'.
    --tconfig=<config-file> JSON config file containing the key word arguments to the ``Tracker`` class (see API) to be
optionally used instead of the corresponding commandline options. If any commandline options are set, they will
override the corresponding arguments provided by the config file.
    --st=<sleep-time>       The number of seconds to sleep in between usage-collection iterations.
    --ru=<ram-unit>         One of 'bytes', 'kilobytes', 'megabytes', 'gigabytes', or 'terabytes'.
    --gru=<gpu-ram-unit>    One of 'bytes', 'kilobytes', 'megabytes', 'gigabytes', or 'terabytes'.
    --tu=<time-unit>        One of 'seconds', 'minutes', 'hours', or 'days'.
    --nec=<num-cores>       The number of cores expected to be used. Defaults to the number of cores in the entire
operating system.
    --guuids=<gpu-uuids>    Comma separated list of the UUIDs of the GPUs for which to track utilization e.g. gpu-
uuid1,gpu-uuid2,etc. Defaults to all the GPUs in the system.
    --disable-logs          If set, warnings are suppressed during tracking. Otherwise, the Tracker logs warnings as usual.
```

```
    --gb=<gpu-brand>        The brand of GPU to profile. Valid values are nvidia and amd. Defaults to the brand of GPU
detected in the system, checking NVIDIA first.
    --tf=<tracking-file>    If specified, stores the individual resource usage measurements at each iteration. Valid file
formats are CSV (.csv) and SQLite (.sqlite) where the SQLite file format stores the data in a table called "data" and
allows for more efficient querying.
    --overwrite             Whether to overwrite the tracking file if it already existed before the beginning of this tracking
session. Do not set if the data in the existing tracking file is still needed.
    sub-track               Perform sub-tracking related commands.
    combine                 Combines multiple sub-tracking files into one. This is usually a result of sub-tracking a code
block that is called in multiple simultaneous processes.
    --stf=<sub-track-file>  The path to the sub-tracking file used to specify the timestamps of specific code-blocks. If not
generated by the gpu-tracker API, must be either a CSV or SQLite file (where the SQLite file contains a table called
"data") where the headers are precisely process_id, code_block_name, position, and timestamp. The process_id is
the ID of the process where the code block is called. code_block_name is the name of the code block. position is
whether it is the start or the stopping point of the code block where 0 represents start and 1 represents stop. And
timestamp is the timestamp where the code block starts or where it stops.
    -p <file-path>          Paths to the sub-tracking files to combine. Must all be the same file format and the same file
format as the resulting sub-tracking file (either .csv or .sqlite). If only one path is provided, it is interpreted as a path to
a directory and all the files in this directory are combined.
    analyze                 Generate the sub-tracking analysis report using the tracking file and sub-tracking file for
resource usage of specific code blocks.
    compare                 Compares multiple tracking sessions to determine differences in computational resource usage
by loading sub-tracking results given their file paths. Sub-tracking results files must be in pickle format e.g. running
the ``sub-track analyze`` command and specifying a file path for ``--output`` and 'pickle' for the ``--format`` option. If
code block results are not included in the sub-tracking files (i.e. no code blocks were sub-tracked), then only overall
results are compared.
    --cconfig=<config-file> JSON config file containing the ``file_path_map`` argument for the ``TrackerComparison``
class and ``statistic`` argument for its ``compare`` method (see API) that can be used instead of the corresponding ``-
m <name>=<path>`` and ``--stat=<statistic>`` commandline options respectively. If additional ``-m <name>=<path>``
options are added on the commandline in addition to a config file, they will be added to the ``file_path_map`` in the
config file. If a ``--stat`` option is provided on the commandline, it will override the ``statistic`` in the config file.
    -m <name>=<file-path>   Mapping of tracking session names to the path of the file containing the sub-tracking
results of said tracking session. Must be in pickle format.
    --stat=<statistic>      The summary statistic of the measurements to compare. One of 'min', 'max', 'mean', or 'std'.
Defaults to 'mean'.
```

The **-e** or **–execute** is a required option where the desired shell command is provided, with both the command and its proceeding arguments surrounded by quotes. Below is an example of running the **bash** command with an argument of **example-script.sh**. When the command completes, its status code is reported.

```
$ gpu-tracker -e "bash example-script.sh" --st=0.3

Resource tracking complete. Process completed with status code: 0
Max RAM:
  Unit: gigabytes
  System capacity: 67.254
  System: 5.61
  Main:
    Total RSS: 0.003
    Private RSS: 0.0
```

```
    Shared RSS: 0.003
  Descendants:
    Total RSS: 0.879
    Private RSS: 0.76
    Shared RSS: 0.119
  Combined:
    Total RSS: 0.881
    Private RSS: 0.761
    Shared RSS: 0.12
Max GPU RAM:
  Unit: gigabytes
  System capacity: 16.376
  System: 1.043
  Main: 0.0
  Descendants: 0.314
  Combined: 0.314
CPU utilization:
  System core count: 12
  Number of expected cores: 12
  System:
    Max sum percent: 324.8
    Max hardware percent: 27.067
    Mean sum percent: 152.109
    Mean hardware percent: 12.676
  Main:
    Max sum percent: 0.0
    Max hardware percent: 0.0
    Mean sum percent: 0.0
    Mean hardware percent: 0.0
  Descendants:
    Max sum percent: 201.8
    Max hardware percent: 16.817
    Mean sum percent: 102.245
    Mean hardware percent: 8.52
  Combined:
    Max sum percent: 201.8
    Max hardware percent: 16.817
    Mean sum percent: 102.245
    Mean hardware percent: 8.52
  Main number of threads: 1
  Descendants number of threads: 12
  Combined number of threads: 13
GPU utilization:
  System GPU count: 1
  Number of expected GPUs: 1
  GPU percentages:
    Max sum percent: 0.0
    Max hardware percent: 0.0
    Mean sum percent: 0.0
    Mean hardware percent: 0.0
Compute time:
  Unit: hours
  Time: 0.001
```

Notice that the RAM and GPU RAM usage primarily takes place in the descendant processes since, in this example, the bash command itself calls the commands relevant to resource usage.

#### 3.1.2.1.2 Options

The units of the computational resources can be modified. For example, **`--tu`** stands for time-unit, **`--gru`** stands for gpu-ram-unit, and **`--ru`** stands for ram-unit.

```
$ gpu-tracker -e 'bash example-script.sh' --tu=seconds --gru=megabytes --ru=megabytes --st=0.2

Resource tracking complete. Process completed with status code: 0
Max RAM:
   Unit: megabytes
   System capacity: 67254.17
   System: 2420.457
   Main:
      Total RSS: 3.109
      Private RSS: 0.319
      Shared RSS: 2.789
   Descendants:
      Total RSS: 849.125
      Private RSS: 731.435
      Shared RSS: 118.125
   Combined:
      Total RSS: 850.338
      Private RSS: 731.754
      Shared RSS: 119.017
Max GPU RAM:
   Unit: megabytes
   System capacity: 16376.0
   System: 1235.0
   Main: 0.0
   Descendants: 506.0
   Combined: 506.0
CPU utilization:
   System core count: 12
   Number of expected cores: 12
   System:
      Max sum percent: 316.4
      Max hardware percent: 26.367
      Mean sum percent: 168.077
      Mean hardware percent: 14.006
   Main:
      Max sum percent: 0.0
      Max hardware percent: 0.0
      Mean sum percent: 0.0
      Mean hardware percent: 0.0
   Descendants:
      Max sum percent: 517.3
      Max hardware percent: 43.108
      Mean sum percent: 130.623
```

```
    Mean hardware percent: 10.885
  Combined:
    Max sum percent: 517.3
    Max hardware percent: 43.108
    Mean sum percent: 130.623
    Mean hardware percent: 10.885
  Main number of threads: 1
  Descendants number of threads: 12
  Combined number of threads: 13
GPU utilization:
  System GPU count: 1
  Number of expected GPUs: 1
  GPU percentages:
    Max sum percent: 5.0
    Max hardware percent: 5.0
    Mean sum percent: 0.462
    Mean hardware percent: 0.462
Compute time:
  Unit: seconds
  Time: 3.995
```

By default, the computational-resource-usage statistics are printed to the screen. The **`-o`** or **`–output`** option can be specified to store that same content in a file.

```
$ gpu-tracker -e 'bash example-script.sh' -o out.txt --st=0.2

Resource tracking complete. Process completed with status code: 0
$ cat out.txt
Max RAM:
  Unit: gigabytes
  System capacity: 67.254
  System: 2.43
  Main:
    Total RSS: 0.003
    Private RSS: 0.0
    Shared RSS: 0.003
  Descendants:
    Total RSS: 0.884
    Private RSS: 0.766
    Shared RSS: 0.118
  Combined:
    Total RSS: 0.885
    Private RSS: 0.766
    Shared RSS: 0.119
Max GPU RAM:
  Unit: gigabytes
  System capacity: 16.376
  System: 1.043
  Main: 0.0
  Descendants: 0.314
  Combined: 0.314
```

```
CPU utilization:
  System core count: 12
  Number of expected cores: 12
  System:
    Max sum percent: 405.0
    Max hardware percent: 33.75
    Mean sum percent: 165.357
    Mean hardware percent: 13.78
  Main:
    Max sum percent: 0.0
    Max hardware percent: 0.0
    Mean sum percent: 0.0
    Mean hardware percent: 0.0
  Descendants:
    Max sum percent: 573.7
    Max hardware percent: 47.808
    Mean sum percent: 124.871
    Mean hardware percent: 10.406
  Combined:
    Max sum percent: 573.7
    Max hardware percent: 47.808
    Mean sum percent: 124.871
    Mean hardware percent: 10.406
  Main number of threads: 1
  Descendants number of threads: 12
  Combined number of threads: 13
GPU utilization:
  System GPU count: 1
  Number of expected GPUs: 1
  GPU percentages:
    Max sum percent: 5.0
    Max hardware percent: 5.0
    Mean sum percent: 0.357
    Mean hardware percent: 0.357
Compute time:
  Unit: hours
  Time: 0.001
```

By default, the format of the output is "text". The **`–f`** or **`–format`** option can specify the format to be "json" instead (or "pickle").

```
$ gpu-tracker -e 'bash example-script.sh' -f json --st=0.2

Resource tracking complete. Process completed with status code: 0
{
 "max_ram": {
  "unit": "gigabytes",
  "system_capacity": 67.2541696,
  "system": 2.5132195840000002,
  "main": {
   "total_rss": 0.00311296,
```

"private_rss": 0.000323584,
"shared_rss": 0.002789376
},
"descendants": {
"total_rss": 0.8446238720000001,
"private_rss": 0.7268597760000001,
"shared_rss": 0.11776409600000001
},
"combined": {
"total_rss": 0.8458403840000001,
"private_rss": 0.7271833600000001,
"shared_rss": 0.11865702400000001
}
},
"max_gpu_ram": {
"unit": "gigabytes",
"system_capacity": 16.376,
"system": 1.235,
"main": 0.0,
"descendants": 0.506,
"combined": 0.506
},
"cpu_utilization": {
"system_core_count": 12,
"n_expected_cores": 12,
"system": {
"max_sum_percent": 316.3,
"max_hardware_percent": 26.358333333333334,
"mean_sum_percent": 167.90769230769232,
"mean_hardware_percent": 13.992307692307692
},
"main": {
"max_sum_percent": 0.0,
"max_hardware_percent": 0.0,
"mean_sum_percent": 0.0,
"mean_hardware_percent": 0.0
},
"descendants": {
"max_sum_percent": 527.1,
"max_hardware_percent": 43.925000000000004,
"mean_sum_percent": 130.81538461538463,
"mean_hardware_percent": 10.90128205128205
},
"combined": {
"max_sum_percent": 527.1,
"max_hardware_percent": 43.925000000000004,
"mean_sum_percent": 130.81538461538463,
"mean_hardware_percent": 10.90128205128205
},
"main_n_threads": 1,
"descendants_n_threads": 12,
"combined_n_threads": 13
},

```
 "gpu_utilization": {
  "system_gpu_count": 1,
  "n_expected_gpus": 1,
  "gpu_percentages": {
   "max_sum_percent": 5.0,
   "max_hardware_percent": 5.0,
   "mean_sum_percent": 0.38461538461538464,
   "mean_hardware_percent": 0.38461538461538464
  }
 },
 "compute_time": {
  "unit": "hours",
  "time": 0.0010899075534608628
 }
}

$ gpu-tracker -e 'bash example-script.sh' -f json -o out.json --st=0.3

Resource tracking complete. Process completed with status code: 0

$ cat out.json

{
 "max_ram": {
  "unit": "gigabytes",
  "system_capacity": 67.2541696,
  "system": 2.325712896,
  "main": {
   "total_rss": 0.0031088640000000002,
   "private_rss": 0.00031948800000000004,
   "shared_rss": 0.002789376
  },
  "descendants": {
   "total_rss": 0.822874112,
   "private_rss": 0.705110016,
   "shared_rss": 0.11776409600000001
  },
  "combined": {
   "total_rss": 0.824086528,
   "private_rss": 0.705429504,
   "shared_rss": 0.11865702400000001
  }
 },
 "max_gpu_ram": {
  "unit": "gigabytes",
  "system_capacity": 16.376,
  "system": 1.235,
  "main": 0.0,
  "descendants": 0.392,
  "combined": 0.392
 },
 "cpu_utilization": {
  "system_core_count": 12,
```

```
  "n_expected_cores": 12,
  "system": {
   "max_sum_percent": 332.1,
   "max_hardware_percent": 27.675,
   "mean_sum_percent": 166.07,
   "mean_hardware_percent": 13.839166666666666
  },
  "main": {
   "max_sum_percent": 0.0,
   "max_hardware_percent": 0.0,
   "mean_sum_percent": 0.0,
   "mean_hardware_percent": 0.0
  },
  "descendants": {
   "max_sum_percent": 104.1,
   "max_hardware_percent": 8.674999999999999,
   "mean_sum_percent": 99.77000000000001,
   "mean_hardware_percent": 8.314166666666665
  },
  "combined": {
   "max_sum_percent": 104.1,
   "max_hardware_percent": 8.674999999999999,
   "mean_sum_percent": 99.77000000000001,
   "mean_hardware_percent": 8.314166666666665
  },
  "main_n_threads": 1,
  "descendants_n_threads": 12,
  "combined_n_threads": 13
 },
 "gpu_utilization": {
  "system_gpu_count": 1,
  "n_expected_gpus": 1,
  "gpu_percentages": {
   "max_sum_percent": 5.0,
   "max_hardware_percent": 5.0,
   "mean_sum_percent": 0.5,
   "mean_hardware_percent": 0.5
  }
 },
 "compute_time": {
  "unit": "hours",
  "time": 0.0010636144214206272
 }
}
```

Alternative to typing out the tracking configuration via command-line options, one can specify a config JSON file via the **`--tconfig`** option.

```
$ cat config.json

{
```

```
  "sleep_time": 0.5,
  "ram_unit": "megabytes",
  "gpu_ram_unit": "megabytes",
  "time_unit": "seconds"
}

$ gpu-tracker -e 'bash example-script.sh' --tconfig=config.json

Resource tracking complete. Process completed with status code: 0
Max RAM:
   Unit: megabytes
   System capacity: 67254.166
   System: 4511.437
   Main:
      Total RSS: 2.957
      Private RSS: 0.319
      Shared RSS: 2.638
   Descendants:
      Total RSS: 894.923
      Private RSS: 781.222
      Shared RSS: 113.701
   Combined:
      Total RSS: 896.135
      Private RSS: 781.541
      Shared RSS: 114.594
Max GPU RAM:
   Unit: megabytes
   System capacity: 16376.0
   System: 727.0
   Main: 0.0
   Descendants: 314.0
   Combined: 314.0
CPU utilization:
   System core count: 12
   Number of expected cores: 12
   System:
      Max sum percent: 259.3
      Max hardware percent: 21.608
      Mean sum percent: 160.9
      Mean hardware percent: 13.408
   Main:
      Max sum percent: 0.0
      Max hardware percent: 0.0
      Mean sum percent: 0.0
      Mean hardware percent: 0.0
   Descendants:
      Max sum percent: 102.8
      Max hardware percent: 8.567
      Mean sum percent: 96.529
      Mean hardware percent: 8.044
   Combined:
      Max sum percent: 102.8
      Max hardware percent: 8.567
```

```
    Mean sum percent: 96.529
    Mean hardware percent: 8.044
  Main number of threads: 1
  Descendants number of threads: 12
  Combined number of threads: 13
GPU utilization:
  System GPU count: 1
  Number of expected GPUs: 1
  GPU percentages:
    Max sum percent: 0.0
    Max hardware percent: 0.0
    Mean sum percent: 0.0
    Mean hardware percent: 0.0
Compute time:
  Unit: seconds
  Time: 3.913
```

##### *3.1.2.2 Sub-tracking*

###### 3.1.2.2.1 Basics

The **`sub-track`** subcommand introduces functionality related to sub-tracking i.e. analyzing computational resource usage for individual code blocks rather than the entire process. This requires a tracking file and a sub-tracking file. The tracking file can be created by specifying the **`--tf`** option when profiling a process using **`--execute`**. The sub-tracking file can be created using the gpu-tracker API i.e. the **`SubTracker`** class. If the process being profiled is not a python script, the sub-tracking file can be generated in any programming language as long as it follows the following format:

It is either a CSV or SQLite file where the headers are **`process_id,code_block_name,position,timestamp`**. The **`process_id`** column is the ID (integer) of the process where the code block was called. The **`code_block_name`** is the given name (string) of the code block to distinguish it from other code blocks being sub-tracked. The **`position`** is an integer of either the value 0 or 1 where 0 indicates the start of the code block and 1 indicates the stopping point of the code block. Finally **`timestamp`** (float) is the timestamp when the code block either starts (where **`position`** is 0) or when it stops (where **`position`** is 1). Both a start timestamp and stop timestamp must be logged for every execution of the code block of interest. If using an SQLite file for more efficient querying of longer tracking sessions, the name of the table must be 'data'.

If sub-tracking a code block that is called in multiple processes, the sub-tracking files of that code block must be unique to each process. For convenience, the **`sub-track combine`** subcommand allows for combining these into a single sub-tracking file that can be used for downstream analysis. This example combines 'sub-tracking1.csv' and 'sub-tracking2.csv' into a single sub-tracking file of the name 'combined-file.csv'. Alternatively, if the **-p** option is only used once, rather than being interpreted as a list of files, it is instead interpreted as the path to a directory containing the sub-tracking files to combine.

```
$ gpu-tracker sub-track combine --stf=combined-file.csv -p sub-tracking1.csv -p sub-tracking2.csv
```

#### 3.1.2.2.2 Analysis

Once a tracking and sub-tracking file is available, the **sub-track analyze** subcommand can generate the sub-tracking results. These can be stored in JSON, text, or pickle format where the pickle format is the same as the **SubTrackingResults** object from the API. If the **--output** option is specified, the content can be stored in the given file path. By default, the content prints to the screen and it is in text format by default.

```
$ gpu-tracker sub-track analyze --tf=tracking.csv --stf=sub-tracking.csv

Overall:
                                              min         max        mean        std
    main_ram                                0.341860    0.944374    0.856037   0.125014
    descendants_ram                         0.000000    0.000000    0.000000   0.000000
    combined_ram                            0.341860    0.944374    0.856037   0.125014
    system_ram                              4.859711    5.553644    5.253445   0.134081
    main_gpu_ram                            0.000000    0.506000    0.429920   0.170432
    descendants_gpu_ram                     0.000000    0.000000    0.000000   0.000000
    combined_gpu_ram                        0.000000    0.506000    0.429920   0.170432
    system_gpu_ram                          0.215000    0.727000    0.650320   0.172010
    gpu_sum_utilization_percent             0.000000    3.000000    0.120000   0.600000
    gpu_hardware_utilization_percent        0.000000    3.000000    0.120000   0.600000
    main_n_threads                         12.000000   15.000000   14.720000   0.842615
    descendants_n_threads                   0.000000    0.000000    0.000000   0.000000
    combined_n_threads                     12.000000   15.000000   14.720000   0.842615
    cpu_system_sum_utilization_percent     11.900000  133.400000  119.212000  22.741909
    cpu_system_hardware_utilization_percent    0.991667   11.116667    9.934333   1.895159
    cpu_main_sum_utilization_percent       78.000000  103.200000   96.924000   6.390767
    cpu_main_hardware_utilization_percent   6.500000    8.600000    8.077000   0.532564
    cpu_descendants_sum_utilization_percent    0.000000    0.000000    0.000000   0.000000
    cpu_descendants_hardware_utilization_percent  0.000000    0.000000    0.000000   0.000000
    cpu_combined_sum_utilization_percent   78.000000  103.200000   96.924000   6.390767
    cpu_combined_hardware_utilization_percent    6.500000    8.600000    8.077000   0.532564
Static Data:
      ram_unit gpu_ram_unit time_unit ram_system_capacity gpu_ram_system_capacity system_core_count
n_expected_cores system_gpu_count n_expected_gpus
     gigabytes    gigabytes     hours          67.254166              16.376             12            12             1            1
Code Block Results:
    Name:                my-code-block
    Num Timepoints:      12
    Num Calls:           3
    Num Non Empty Calls: 3
    Compute Time:
                  min       max      mean       std     total
          2.580433  2.789909  2.651185  0.120147  7.953554
    Resource Usage:
                                              min         max        mean        std
```

```
    main_ram                                    0.341860    0.936559    0.808736   0.167663
    descendants_ram                             0.000000    0.000000    0.000000   0.000000
    combined_ram                                0.341860    0.936559    0.808736   0.167663
    system_ram                                  4.859711    5.553644    5.231854   0.191567
    main_gpu_ram                                0.000000    0.506000    0.363500   0.225892
    descendants_gpu_ram                         0.000000    0.000000    0.000000   0.000000
    combined_gpu_ram                            0.000000    0.506000    0.363500   0.225892
    system_gpu_ram                              0.215000    0.727000    0.583250   0.228088
    gpu_sum_utilization_percent                 0.000000    0.000000    0.000000   0.000000
    gpu_hardware_utilization_percent            0.000000    0.000000    0.000000   0.000000
    main_n_threads                             12.000000   15.000000   14.416667   1.164500
    descendants_n_threads                       0.000000    0.000000    0.000000   0.000000
    combined_n_threads                         12.000000   15.000000   14.416667   1.164500
    cpu_system_sum_utilization_percent         11.900000  130.800000  113.641667  32.352363
    cpu_system_hardware_utilization_percent     0.991667   10.900000    9.470139   2.696030
    cpu_main_sum_utilization_percent           79.600000  103.100000   96.583333   6.726587
    cpu_main_hardware_utilization_percent       6.633333    8.591667    8.048611   0.560549
    cpu_descendants_sum_utilization_percent     0.000000    0.000000    0.000000   0.000000
    cpu_descendants_hardware_utilization_percent  0.000000    0.000000    0.000000   0.000000
    cpu_combined_sum_utilization_percent       79.600000  103.100000   96.583333   6.726587
    cpu_combined_hardware_utilization_percent   6.633333    8.591667    8.048611   0.560549

Name:              my-function
Num Timepoints:      12
Num Calls:           3
Num Non Empty Calls: 3
Compute Time:
          min       max      mean       std     total
    2.538011  2.577679  2.553176  0.021419  7.659528
Resource Usage:
                                                 min         max        mean       std
    main_ram                                    0.864592    0.944374    0.896998  0.034505
    descendants_ram                             0.000000    0.000000    0.000000  0.000000
    combined_ram                                0.864592    0.944374    0.896998  0.034505
    system_ram                                  5.203415    5.315219    5.271566  0.038751
    main_gpu_ram                                0.314000    0.506000    0.490000  0.055426
    descendants_gpu_ram                         0.000000    0.000000    0.000000  0.000000
    combined_gpu_ram                            0.314000    0.506000    0.490000  0.055426
    system_gpu_ram                              0.535000    0.727000    0.711000  0.055426
    gpu_sum_utilization_percent                 0.000000    3.000000    0.250000  0.866025
    gpu_hardware_utilization_percent            0.000000    3.000000    0.250000  0.866025
    main_n_threads                             15.000000   15.000000   15.000000  0.000000
    descendants_n_threads                       0.000000    0.000000    0.000000  0.000000
    combined_n_threads                         15.000000   15.000000   15.000000  0.000000
    cpu_system_sum_utilization_percent        120.300000  133.400000  124.566667  4.001439
    cpu_system_hardware_utilization_percent    10.025000   11.116667   10.380556  0.333453
    cpu_main_sum_utilization_percent           94.700000  103.200000   98.841667  2.677332
    cpu_main_hardware_utilization_percent       7.891667    8.600000    8.236806  0.223111
    cpu_descendants_sum_utilization_percent     0.000000    0.000000    0.000000  0.000000
    cpu_descendants_hardware_utilization_percent  0.000000    0.000000    0.000000  0.000000
    cpu_combined_sum_utilization_percent       94.700000  103.200000   98.841667  2.677332
    cpu_combined_hardware_utilization_percent   7.891667    8.600000    8.236806  0.223111
```

The overall resource usage of the tracking session is provided as well as its static data. This is followed by the compute time and resource usage of each code block.

#### 3.1.2.2.3 Comparison

Storing the results of the sub-tracking analysis in a pickle file allows for one tracking session to be compared to another.

```
$ gpu-tracker sub-track analyze --tf=tracking.csv --stf=sub-tracking.csv --format=pickle --output=my-results.pkl
```

The **`sub-track compare`** subcommand compares the computational resource usage of multiple tracking sessions. This is useful when you want to determine how a change can impact the computational efficiency of your process, whether it be different input data, an alternative implementation, etc. The **`-m`** option creates a mapping from the given name of a tracking session to the file path where its sub-tracking results are stored in pickle format. Say you wanted to call one tracking session 'A' and then the second tracking session 'B' where the results of tracking session 'A' are stored in 'results.pkl' and that of session 'B' are in 'results2.pkl'.

```
$ gpu-tracker sub-track compare -m A=results.pkl -m B=results2.pkl

Overall Resource Usage:
  Main Ram:
               B         A
       0.856037  0.861921
  Descendants Ram:
         A    B
       0.0  0.0
  Combined Ram:
               B         A
       0.856037  0.861921
  System Ram:
               B         A
       5.253445  5.281926
  Main Gpu Ram:
              B         A
       0.42992  0.448364
  Descendants Gpu Ram:
         A    B
       0.0  0.0
  Combined Gpu Ram:
              B         A
       0.42992  0.448364
  System Gpu Ram:
              B         A
       0.65032  0.668909
  Gpu Sum Utilization Percent:
         A     B
```

```
        0.0  0.12
    Gpu Hardware Utilization Percent:
          A     B
        0.0  0.12
    Main N Threads:
              B          A
        14.72  14.757576
    Descendants N Threads:
          A    B
        0.0  0.0
    Combined N Threads:
              B          A
        14.72  14.757576
    Cpu System Sum Utilization Percent:
                B           A
        119.212  121.918182
    Cpu System Hardware Utilization Percent:
                 B          A
        9.934333  10.159848
    Cpu Main Sum Utilization Percent:
               B          A
        96.924  99.060606
    Cpu Main Hardware Utilization Percent:
              B         A
        8.077  8.255051
    Cpu Descendants Sum Utilization Percent:
          A    B
        0.0  0.0
    Cpu Descendants Hardware Utilization Percent:
          A    B
        0.0  0.0
    Cpu Combined Sum Utilization Percent:
               B          A
        96.924  99.060606
    Cpu Combined Hardware Utilization Percent:
              B         A
        8.077  8.255051
Code Block Resource Usage:
    Main Ram:
my-code-block:
                        B         A
                0.808736  0.846999
my-function:
                        A         B
                0.888034  0.896998
    Descendants Ram:
my-code-block:
                  A    B
                0.0  0.0
my-function:
                  A    B
                0.0  0.0
    Combined Ram:
```

```
my-code-block:
                B         A
            0.808736  0.846999
my-function:
                A         B
            0.888034  0.896998
    System Ram:
my-code-block:
                A         B
            5.170665  5.231854
my-function:
                B         A
            5.271566  5.476632
    Main Gpu Ram:
my-code-block:
              B         A
            0.3635  0.415429
my-function:
             B      A
            0.49  0.506
    Descendants Gpu Ram:
my-code-block:
             A    B
            0.0  0.0
my-function:
             A    B
            0.0  0.0
    Combined Gpu Ram:
my-code-block:
              B         A
            0.3635  0.415429
my-function:
             B      A
            0.49  0.506
    System Gpu Ram:
my-code-block:
               B         A
            0.58325  0.635714
my-function:
              B      A
            0.711  0.727
    Gpu Sum Utilization Percent:
my-code-block:
             A    B
            0.0  0.0
my-function:
             A     B
            0.0  0.25
    Gpu Hardware Utilization Percent:
my-code-block:
             A    B
            0.0  0.0
my-function:
```

```
         A     B
       0.0  0.25
    Main N Threads:
my-code-block:
              B          A
       14.416667  14.619048
my-function:
          A     B
       15.0  15.0
    Descendants N Threads:
my-code-block:
         A    B
       0.0  0.0
my-function:
         A    B
       0.0  0.0
    Combined N Threads:
my-code-block:
              B          A
       14.416667  14.619048
my-function:
          A     B
       15.0  15.0
    Cpu System Sum Utilization Percent:
my-code-block:
               B           A
       113.641667  120.142857
my-function:
               B        A
       124.566667  125.025
    Cpu System Hardware Utilization Percent:
my-code-block:
             B          A
       9.470139  10.011905
my-function:
              B         A
       10.380556  10.41875
    Cpu Main Sum Utilization Percent:
my-code-block:
              B          A
       96.583333  98.652381
my-function:
              B       A
       98.841667  99.775
    Cpu Main Hardware Utilization Percent:
my-code-block:
             B         A
       8.048611  8.221032
my-function:
             B         A
       8.236806  8.314583
    Cpu Descendants Sum Utilization Percent:
my-code-block:
```

```
                A    B
              0.0  0.0
my-function:
                A    B
              0.0  0.0
   Cpu Descendants Hardware Utilization Percent:
my-code-block:
                A    B
              0.0  0.0
my-function:
                A    B
              0.0  0.0
   Cpu Combined Sum Utilization Percent:
my-code-block:
                    B          A
              96.583333  98.652381
my-function:
                    B        A
              98.841667  99.775
   Cpu Combined Hardware Utilization Percent:
my-code-block:
                   B         A
              8.048611  8.221032
my-function:
                   B         A
              8.236806  8.314583
Code Block Compute Time:
my-code-block:
              B        A
         2.651185  2.68558
my-function:
              B         A
         2.553176  2.559218
```

Both the overall usage is compared and per code block. The default format is text and the default output is printing to the console. The **`--format`** and **`--output`** options can be configured similarly to those in the **`sub-track analyze`** subcommand. By default, the 'mean' of measurements is compared. Alternatively, the **`--stat`** option can be set to 'min', 'max', or 'std' to compare a different summary statistic.

## 3.2 Profiling Example 1

To provide a real-world use case of gpu-tracker, Huckvale and Moseley used the package in their work involving training machine learning models (Erik D Huckvale and Moseley, 2024c) (while a prototype of gpu-tracker was used in the following machine learning-focused publications: Huckvale and Moseley (Erik D Huckvale and Moseley, 2024a); Huckvale et al. (Huckvale et al., 2023)). It's a common need to profile the resource usage of model training and evaluation to compare different models, datasets, and implementations. Huckvale and Moseley trained two types of models, namely XGBoost (Chen and Guestrin, 2016) and multi-layer-perceptron (MLP) (Bisong, 2019). And they used two variants of the dataset, i.e. one with the full number of features (variables) and a compressed version, where a compression method was used to map the original features to a smaller number of extracted features. The profiling results of their analyses are shown in Table 1. For rigorous comparisons , machine learning models are frequently evaluated over multiple iterations and the usage measurements of Table 1 are the results of 50 iterations. We see that despite the compressed data having less features, the MLP required more compute time to train on it.

*Table 1 - Computational resource usage of different machine learning models and datasets.*

| Model | Computational Resource Type | Value |
|---|---|---|
| MLP | RAM (gigabytes) | 2.507 |
| | GPU RAM (gigabytes) | 0.848 |
| | Compute time (minutes) | 89.706 |
| MLP with Encoding | RAM (gigabytes) | 1.764 |
| | GPU RAM (gigabytes) | 0.42 |
| | Compute time (minutes) | 129.130 |
| XGBoost | RAM (gigabytes) | 23.878 |
| | GPU RAM (gigabytes) | 11.052 |
| | Compute time (minutes) | 70.098 |
| XGBoost with Encoding | RAM (gigabytes) | 5.307 |
| | GPU RAM (gigabytes) | 5.776 |
| | Compute time (minutes) | 27.297 |

Figure 4 shows the code snippet, the SLURM (Yoo et al., 2003) submission scripts, and terminal commands demonstrating how gpu-tracker was used in an HPC environment to profile the training of the machine learning models described in Table 1.

(a) Code snippet using gpu-tracker API to profile model training.

```
with gput.Tracker(time_unit='minutes') as tracker:
    for i, train_test_split in enumerate(tqdm.tqdm(self._cross_validation(
            n_iterations=ModelTrainer.TRAIN_AND_EVALUATE_N_CV_ITERATIONS,
test_size=ModelTrainer.TEST_SIZE, **hyperparameters),
            total=ModelTrainer.TRAIN_AND_EVALUATE_N_CV_ITERATIONS)):
        result = self._fit_and_predict(train_test_split, **hyperparameters)
...
```

(b) SLURM job submission script to train the XGBoost models (train-mpp-xgboost.sh).

```
sbatch <<EOT
#!/bin/bash
#SBATCH --time=72:00:00
#SBATCH --job-name="train-mpp-xgboost${3}${2}-${1}"
#SBATCH --ntasks=1
```

```
#SBATCH --cpus-per-task=10
#SBATCH --mem-per-cpu=17G
#SBATCH --partition=P4V12_SKY32M192_L
#SBATCH -e "slurm/train-mpp-xgboost${3}${2}-${1}.err"
#SBATCH -o "slurm/train-mpp-xgboost${3}${2}-${1}.out"
#SBATCH -A gol_hmo222_uksr
#SBATCH --gpus 1
#SBATCH --gres=gpu

python3 train_test_mpp.py xgboost ${2} --nt=50 --nw=9 --bs=${1} ${3}
EOT
```

(c) SLURM job submission script to train the MLP models (train-mpp-mlp.sh).

```
sbatch <<EOT
#!/bin/bash
#SBATCH --time=72:00:00
#SBATCH --job-name="train-mpp-mlp${3}${2}-${1}"
#SBATCH --ntasks=1
#SBATCH --cpus-per-task=16
#SBATCH --mem-per-cpu=6G
#SBATCH --partition=P4V12_SKY32M192_L
#SBATCH -e "slurm/train-mpp-mlp${3}${2}-${1}.err"
#SBATCH -o "slurm/train-mpp-mlp${3}${2}-${1}.out"
#SBATCH -A gol_hmo222_uksr
#SBATCH --gpus 1
#SBATCH --gres=gpu

python3 train_test_mpp.py mlp ${2} --nt=50 --nw=15 --bs=${1} ${3}
EOT
```

(d) Terminal commands for running the job submission scripts.

```
bash train-mpp-mlp.sh 1024 --level=2
bash train-mpp-mlp.sh 1024 --level=2 --encoded
bash train-mpp-xgboost.sh 100000 --level=2
bash train-mpp-xgboost.sh 100000 --level=2 --encoded
```

*Figure 4 – Code snippet within a script using the gpu-tracker API to profile resources used when training various machine learning models / datasets along with the SLURM job submission scripts for running this Python script and the terminal commands for running the job submission scripts.*

## 3.3 Profiling Example 2

While the first profiling example compares different types of machine learning models, the example in Table 2 just compares different datasets. The publication using gpu-tracker in Table 2 (Erik D Huckvale and Moseley, 2024b) had two machine learning datasets: ‘Combined’ and ‘L3 Only’. These two datasets were both scaled and not scaled, resulting in four datasets total. The compute time and RAM (CPU and GPU) are compared between the different datasets.

*Table 2 - Profiling example comparing the computational resource usage of different machine learning datasets.*

| Dataset | Scaled | Resource | Unit | Amount |
|---|---|---|---|---|
| Combined | True | Compute time | Hours | 131.8 |
| | | GPU RAM | Gigabytes | 4.3 |
| | | RAM | Gigabytes | 13.6 |
| | False | Compute time | Hours | 131.1 |
| | | GPU RAM | Gigabytes | 4.6 |
| | | RAM | Gigabytes | 14.1 |
| L3 only | True | Compute time | Hours | 171.3 |
| | | GPU RAM | Gigabytes | 2.9 |
| | | RAM | Gigabytes | 12.6 |
| | False | Compute time | Hours | 127.8 |
| | | GPU RAM | Gigabytes | 4.1 |
| | | RAM | Gigabytes | 16.6 |

Figure 5 shows the code snippet, the SLURM (Yoo et al., 2003) submission scripts, and terminal commands demonstrating how gpu-tracker was used in an HPC environment to profile the training of an MLP on various datasets described in Table 2.

(a) Code snippet using the gpu-tracker API to profile model training.

```
with gput.Tracker() as tracker:
    for i, train_test_split in enumerate(tqdm.tqdm(self._cross_validation(
            n_iterations=ModelTrainer.TRAIN_AND_EVALUATE_N_CV_ITERATIONS,
test_size=ModelTrainer.TEST_SIZE,
            start_iteration=start_iteration, save_random_state=True, **hyperparameters),
            total=ModelTrainer.TRAIN_AND_EVALUATE_N_CV_ITERATIONS - start_iteration)):
        i += start_iteration
        result = self._fit_and_predict(train_test_split, **hyperparameters)
```

(b) SLURM job submission script to train an MLP on various datasets (train-mpp-mlp.sh).

```
sbatch <<EOT
#!/bin/bash
#SBATCH --time=72:00:00
#SBATCH --job-name="train-mpp-mlp${3}${2}-${1}"
#SBATCH --ntasks=1
#SBATCH --cpus-per-task=5
```

```
#SBATCH --mem-per-cpu=12G
#SBATCH --partition=P4V12_SKY32M192_L
#SBATCH -e "slurm/train-mpp-mlp${3}${2}-${1}.err"
#SBATCH -o "slurm/train-mpp-mlp${3}${2}-${1}.out"
#SBATCH -A gol_hmo222_uksr
#SBATCH --gpus 1
#SBATCH --gres=gpu
#SBATCH --mail-type ALL

if [[ -z "${1}" ]]
then
	>&2 echo "A batch size was not provided as the first argument."
	exit 1
fi

source ~/.bashrc
conda activate acmpp
time python3 train_test_mpp.py mlp ${2} --nt=200 --nw=4 --bs=${1} ${3}
EOT
```

(c) Terminal commands for running the job submission script.

```
bash train-mpp-mlp.sh 10000 --scaled
bash train-mpp-mlp.sh 10000
bash train-mpp-mlp.sh 10000 --level=3 --scaled
bash train-mpp-mlp.sh 10000 --level=3
```

*Figure 5 - Code snippet within a script using the gpu-tracker API to profile resources used when training an MLP on various datasets along with the SLURM job submission script for running this Python script and the terminal commands for running the job submission script.*

## 3.4 Profiling Example 3

Machine learning methods often train models on data in batches that are stored in the GPU RAM. As such, there has been work in developing efficient data loading methods. Table 3 compares a prior and new data loading method for training GPU accelerated neural networks (Erik D. Huckvale and Moseley, 2024). We see that the new data loading method increases GPU utilization by over 10-fold with only a 4.2 GB increased usage of GPU RAM. This order of magnitude increase in GPU utilization lowers the training time by over 20-fold. This profiling example exemplifies why GPU profiling is invaluable for efficient utilization of GPU hardware resources. Without gpu-tracker's profiling, the GPU underutilization would not have been detected and remedied. Also, the full analysis in this paper took approximately 17 hours and 42 minutes on an NVIDIA H100. Without the improvement in GPU utilization, the analysis would have taken over 367 hours of execution time.

*Table 3 - Comparing GPU usage and compute time of two different data loading methods for training neural networks.*

| Data Loading Method | Resource | Unit | Amount |
|---|---|---|---|

| Old | CPU Utilization | % | 62.6 |
|---|---|---|---|
| | GPU RAM | Gigabytes | 9.0 |
| | GPU Utilization | % | 8.8 |
| | RAM | Gigabytes | 22.8 |
| | Real Time | Minutes | 978.2 |
| New | CPU Utilization | % | 98.2 |
| | GPU RAM | Gigabytes | 13.2 |
| | GPU Utilization | % | 93.7 |
| | RAM | Gigabytes | 3.9 |
| | Real Time | Minutes | 47.1 |

Figure 6 shows the code snippet within a Python script and the command within a SLURM job submission script to run the Python script to demonstrate how gpu-tracker was used to compare the old and new data loading techniques described in Table 3.

(a) Code snippet from the Python script that uses gpu-tracker to compare the two data loading methods.

```
if tracker is not None:
    tracker.start()
trainer.fit(model, train_dataloaders=train_data_loader)
if tracker is not None:
    tracker.stop()
```

(b) The command within the SLURM job submission script to run the Python script.

```
python3 compare_data_loaders.py
```

*Figure 6 - Code snippet within a Python script using the gpu-tracker API to compare two data loading techniques along with the command used within a SLURM job submission script to run the Python script.*

# 4 Related Work

Table 4 compares the features provided by gpu-tracker to those provided by other tools of a similar purpose. The nvprof package is a similar tool for Nvidia GPUs, but has become deprecated and replaced by Nvidia's nsight commandline tool (Nvidia, 2025a). The nsight package tracks and reports a myriad of different stats related to GPU and CPU usage, including the capability to track individual timepoints like gpu-tracker can. While there is a richness of information included, this comes with a more complex interface and higher learning curve as compared to the relatively simple interface and lightweight implementation and installation of gpu-tracker. The nsight package provides a GUI unlike any of the other tools compared in Table 4; however, it lacks an API such that it can be integrated into specific parts of a user's code. Like gpu-tracker, nsight not only tracks resource usage of the main process but it also tracks resource usage of all child processes. But support for system-wide resource tracking varies depending on the computational resource in question and the operating system. Peak GPU RAM and utilization percentage are reported but the average of these would need to be manually calculated by the user. CPU utilization percentage is only tracked on Linux and CPU RAM is not tracked at all. Additionally, tracking GPU memory increases the overhead of the process(es) being tracked whereas gpu-tracker does not impact the performance of the tracked process, since it simply queries nvidia-smi in a separate process. In addition to profiling an entire application, the nsight package is capable of sub-tracking individual calls to functions or smaller code blocks as is gpu-tracker. However, for nsight to accomplish this feature, it depends on using an extension library known as nvtx (Nvidia, 2025b). The nvtx package can be used to profile function calls and code blocks with an API that's quite similar to the **`SubTracker`** context manager and **`sub_track`** decorator in the **`sub_tracker`** module of gpu-tracker. The nvtx package has an API for C/C++ and Python only and produces a **`.qdrep`** file. This file format is less well known than the standard CSV and SQLite file formats, causing a challenge for developers using programming languages other than C/C++ or Python. However, gpu-tracker only requires a simple sub-tracking file in tabular format and with only 4 columns, enabling users to create their own sub-tracking files from any programming language they use. The nsight package is helpful for those who want in-depth information, have deep knowledge and interest in the intricate inner workings of GPUs, and do not require CPU memory tracking, automatic calculation of average GPU usage, and full support for CPU percentage and system-wide resource tracking across all platforms. The gpu-tracker package meets these additional use cases, but has an easy-to-learn interface, is lightweight, and provides straightforward and automatically calculated output, conveniently including real compute time, without any additional dependencies nor effort on the part of the user.

*Table 4 - Feature comparison between gpu-tracker and similar tools (Blue=Fully supports the feature; Yellow=Partially supports the feature; Red=Does not have the feature).*

| Tool Name | API | CLI | GUI | Max and Mean GPU RAM | Max and Mean CPU RAM | Max and Mean GPU Utilization | Max and Mean CPU Utilization | Child Processes and Entire System | Arbitrary GPU Tracking | Trackin Resources Individu Timepoin |
|---|---|---|---|---|---|---|---|---|---|---|
| gpu-tracker | Yes | Yes | No | Yes | Yes | Yes | Yes | Yes | Yes | Yes |
| nsight | No | Yes | Yes | Max Only | No | Max Only | Only On Linux | Not All Platforms Support System-wide | Yes | Yes |

| mem-tracker | Yes | No | No | Max Only | Max Only | No | No | Main and Children Only | Yes | No |
|---|---|---|---|---|---|---|---|---|---|---|
| pytorch-memory-utils | Yes | No | No | Max Only | No | No | No | Main Process Only | PyTorch Objects Only | No |
| tensorflow-memory-utils | Yes | No | No | Max Only | No | No | No | Main Process Only | TensorFlow Objects Only | No |

Also compared in Table 4 is a package known as memtracker (Lan, 2023) with some similarities to our implementation. These similarities include using the nvidia-smi (Nvidia, 2024) shell command (in a subprocess) to obtain GPU RAM data and using the psutil package (Rodola, 2024) to obtain RAM data. The memtracker package contains a class that uses an underlying thread, rather than a subprocess, with a resource collection loop and also has **`start()`** and **`stop()`** methods that respectively start and stop its profiling thread. They likewise also collect RAM of child processes. The similarities in the memtracker and gpu-tracker tracking class design developed by two independent groups provides solid evidence that that the base single class design is fairly optimal. However, a significant difference between memtracker and gpu-tracker is that memtracker is only intended to profile a specified Python function whereas gpu-tracker can profile an arbitrary block of code, providing the choice of containing the code within a context manager or explicitly calling the **`start()`** and **`stop()`** methods. The **`start()`** and **`stop()`** methods of the API are especially helpful in environments such as Jupyter notebooks (Kluyver et al., 2016) or the Python console where a user may not desire to define a function or complete a code block before they want to profile, but rather can profile interactive code. The sub-tracking functionality of gpu-tracker allows tracking various functions or code blocks without creating an additional sub-process (or thread in the case of memtracker) for each one. Even more notable is the CLI provided by gpu-tracker which removes the restriction of only profiling Python code, being able to track resources of any shell command of any language. In fact, the CLI removes the requirement of the user to write any code of any language at all, expanding the audience to non-software-developers and providing maximal flexibility. While the API of memtracker allows profiling the process in which the specified function is called, the API of gpu-tracker likewise defaults to the current process but can optionally accept an arbitrary process ID. The gpu-tracker package intelligently takes into account shared memory across processes on Linux distributions, distinguishing between private RSS and shared RSS, and sums the overall usage of processes by only counting shared memory once to avoid overestimation. These increases in flexibility and granularity come with increased likelihood of encountering edge cases during execution. This is why gpu-tracker also provides thorough error handling, particularly surrounding edges cases of the subprocess failing to terminate properly and race conditions of processes no longer existing. And helpful warning messages are logged to inform the user of exceptional occurrences and aid them in troubleshooting if necessary. Other features of gpu-tracker that go beyond the basic functionality of memtracker include options for choosing different measurement units (e.g. hours, gigabytes, etc.), industry standard unit testing, and sophisticated DevOps pipelines for running the tests on all three of the major operating systems (Windows, MacOS, and Linux) and for automatically building and web-hosting professional documentation. The documentation is sufficiently detailed, specifying all of the classes, parameters, methods, and attributes of the API and all the options of the CLI, and it provides in-depth tutorials. In addition, the CPU utilization and GPU utilization is also provided by gpu-tracker. It should also be noted that threads in Python are not truly parallel so the use of a subprocess instead of a thread enables tracking resource usage measurements that otherwise would have been missed due to the GIL needing to switch between the main thread and tracking thread. We discovered this limitation with a thread implementation in an earlier prototype of gpu-tracker and quickly selected the superior subprocess implementation for full development.

Finally, PyTorch (Paszke et al., 2019) and TensorFlow (Abadi et al., 2016) are also listed in Table 4. These libraries involve creating and processing tensors usually for the purpose of implementing neural networks as well as performing deep learning. Use of these libraries is primarily accelerated by the GPU. They both provide built-in GPU support and likewise provide GPU profiling tools. The profiling, however, is specific to the tensors and models provided by the libraries themselves and they are not intended to profile the GPU used by any other software. They also are intended to monitor the main process and do not automatically profile child processes nor system-wide GPU usage. If a user is planning on only using the GPU for one of these two libraries and is mostly just concerned with GPU RAM in the main process, then PyTorch or TensorFlow is likely sufficient. They are specific to profiling the use of the library itself, particularly on the GPU, not providing support for profiling other types of hardware, and are not generic profiling tools such as gpu-tracker. Note that these and all other packages compared to gpu-tracker in Table 4 only support Nvidia GPUs whereas gpu-tracker both supports Nvidia and AMD (AMD, 2024) GPUs.

# 5 Conclusion

The analysis of comparing machine learning models and implementations in our profiling examples demonstrates the importance of comprehensive resource tracking. In the field of machine learning, feature reduction generally is motivated by decreasing the training time. However in Profiling Example 1 (Table 1), we see for MLP training that compressing the data actually resulted in increased training time due to other factors beyond the number of features. Objective data collection helps guide our decision making when optimizing computing tasks as our expectations are not always met. We also see that all models required less than 12 gigabytes of GPU RAM, indicating that a more expensive 16-gigabyte or 32-gigabyte GPU was not necessary in our example. This information saves users monetarily as far as purchasing hardware and is also courteous to other users in an HPC system when allocating and sharing computational resources. Similarly, the profiling example in Table 3 shows a huge improvement in GPU utilization percentage between two data loading methods, where the improved data loading method decreases the model training time by over 20-fold. Without the improvement in GPU utilization, the full analysis would have taken over 367 hours of execution time. To put this into a common computational research context, many academic HPC resources have wall times of 24 hours, making multiday executions difficult or near impossible to complete. Also, the increase in training efficiency enables larger data sets to be used, opening doors to new possibilities. As demonstrated, the GPU profiling provided by gpu tracker helps identify bottlenecks and test improvement of methods that alleviate such bottlenecks.

With increasing usage of HPC systems, especially in the rise of machine learning and artificial intelligence applications that make excessive use of GPUs, there is an increasing need for a computational resource profiling tool with GPU support. The increasing demand for computational efficiency results in more use of multiprocessing, necessitating resource profiling for multiple processes in addition to the main process of a job. While there have been previous efforts to profile compute time and RAM, minimal options have been available for profiling GPU RAM. The gpu-tracker package provides a solution for determining peak GPU RAM usage and conveniently tracks motherboard RAM and compute time within the same user-friendly interfaces. We've demonstrated that gpu-tracker seamlessly profiles computing tasks, both the main and descendant processes, with minimum overhead or configuration, both in desktop and HPC environments, providing the most feature-rich and robust computational resource profiling tool to date. Currently, the gpu tracker package provides comprehensive functionality for profiling CPUs, Nvidia GPUs, and AMD GPUs. The API can be used directly in Python code, while the CLI functionality provides language agnostic functionality.

# Acknowledgements

We thank the University of Kentucky Center for Computational Sciences and Information Technology Services Research Computing for their support and use of the Lipscomb Compute Cluster and associated research computing resources.

# Statements and Declarations

## Ethical Considerations

Not applicable.

## Consent to Participate

Not applicable.

## Consent for Publication

Not applicable.

## Declaration of Conflicting Interest

The authors declared no potential conflicts of interest with respect to the research, authorship, and/or publication of this article.

## Funding Statement

This research was funded by the National Science Foundation, grant number 2020026 (PI Moseley), and by the National Institutes of Health, grant number P42 ES007380 (University of Kentucky Superfund Research Program Grant; PI Pennell). The content is solely the responsibility of the authors and does not necessarily represent the official views of the National Science Foundation nor the National Institute of Environmental Health Sciences.

## Data Availability

Code for reproducing the novel results of this manuscript is available at the following Figshare item: https://doi.org/10.6084/m9.figshare.27064810.